\documentclass[a4papert]{article}
\pdfoutput=1
\usepackage[affil-it]{authblk}

\usepackage[latin1]{inputenc}
\usepackage[T1]{fontenc}
\usepackage[english]{babel}
\usepackage{amsfonts}
\usepackage{amssymb}
\usepackage{amsmath}
\usepackage{amsthm}
\usepackage{graphicx,xcolor}
\usepackage{a4wide}
\usepackage{bbm}
\usepackage{psfrag}
\usepackage{color}
\usepackage{subfigure}
\usepackage{amsopn}
\usepackage{url}

\usepackage{esint}

\makeatletter

\@ifundefined{textcolor}{}
{
 \definecolor{BLACK}{gray}{0}
 \definecolor{WHITE}{gray}{1}
 \definecolor{RED}{rgb}{1,0,0}
 \definecolor{GREEN}{rgb}{0,1,0}
 \definecolor{BLUE}{rgb}{0,0,1}
 \definecolor{CYAN}{cmyk}{1,0,0,0}
 \definecolor{MAGENTA}{cmyk}{0,1,0,0}
 \definecolor{YELLOW}{cmyk}{0,0,1,0}
}

\newcommand{\cref}{\ref}

\DeclareMathOperator{\var}{var}
\DeclareMathOperator{\cov}{cov}

\renewcommand{\epsilon}{\varepsilon}

\theoremstyle{nonumberplain}
\newtheorem{oss}{Remarks}      
\newtheorem{proposition}{Proposition}

\makeatother

\begin{document}

\title{Resolution of ranking hierarchies in directed networks}

\author{Elisa Letizia}
 \affil{%
 Scuola Normale Superiore, Piazza dei Cavalieri 7, Pisa, 56126, Italy  }%
\author{Paolo Barucca}%
\affil{%
University of Zurich, Sch\"onberggasse 1, Z\"urich, 8001 Switzerland \\
LIMS, 35a South Street, London, W1K 2XF, United Kingdom }%
\author{Fabrizio Lillo}%
\affil{%
Department of Mathematics, University of Bologna, Piazza di Porta San Donato 5, Bologna, 40126, Italy}%

\maketitle

\begin{abstract}
Identifying hierarchies and rankings of nodes in directed graphs is fundamental in many applications such as social network analysis, biology, economics, and finance. A recently proposed method identifies the hierarchy by finding the ordered partition of nodes which minimises a score function, termed agony. This function penalises the links violating the hierarchy in a way depending on the strength of the violation. 
To investigate the resolution of ranking hierarchies we introduce an ensemble of random graphs, the Ranked Stochastic Block Model. We find that agony may fail to identify hierarchies when the structure is not strong enough and the size of the classes is small with respect to the whole network. We analytically characterise the resolution threshold and we show that an iterated version of agony can partly overcome this resolution limit.
\end{abstract}

\section{Introduction}

Identifying ranking hierarchies in complex networks is of paramount importance in many disciplines and applications. An exact hierarchical organisation in a directed network means that the set of nodes can be divided in an ordered collection of classes such that links exist only from a node of a low rank class to a node of a higher rank class\footnote{Clearly it is equivalent to define exact hierarchical structures when links exist only from upper to lower classes.}. 

Recently the relevance of measuring ranking hierarchy was pointed out in the context of ecosystems \cite{johnson2014trophic}, in which it was shown how species exhibit a property of trophic coherence, measuring how consistently a species falls into a distinct level of hierarchy within a food web. Other major applications include social network analysis \cite{Shetty2005,nguyen2014data}, the study of funds flow in financial networks \cite{fama2002testing,frank2003testing}, and of corporate cross-ownerships in economics \cite{glattfelder2009backbone}.

Since real networks are not necessarily exactly hierarchical, the problem considered here is to find an \emph{optimal} ordered partition of nodes into classes such that the structure has a maximal level of hierarchy. 

Framed in this way, the procedure is to choose a suitable hierarchy metric and to devise feasible algorithms which find the ordered partition of nodes in such a way that the hierarchy metric is maximised. In \cite{tatti2015hierarchies}, this maximisation was recognised to be a dual problem of circulation, known to be related to the cost max-flow minimisation \cite{orlin1993faster}. The problem is analogous to the more explored problem of community detection in graphs \cite{fortunatoReport}. In such case a common approach is  to choose a metric, for example the modularity, and to look for partitions that maximizes it. It is well known that modularity has resolution limits \cite{kumpula2007limited,fortunato2007resolution}, and the associated optimisation problem might be a hard computational task, even if successful heuristics exist \cite{louvain}.

It is important to stress that the concept of {\it ranking hierarchy} we employ in this paper, introduced in \cite{simon1991architecture}, and further developed in \cite{krackhardt1994graph,maiya2009inferring,Gupte2011,tatti2015hierarchies,Romei2015}, models graphs, representing for example social organisations, as command structure or influential communities. 

Related literature sharing a similar definition of hierarchy includes \cite{trusina2004hierarchy, tibely2013extracting, nepusz2013hierarchical, corominas2013origins, mones2013hierarchy}. This concept is therefore very different from the more common definition of {\it nested hierarchy} in networks, studied for example in \cite{Clauset2007,Clauset2008,newman2012communities,peixoto2014hierarchical}, where low-level communities of nodes are nested into bigger ones, in a way directly associated with hierarchical clustering. The former concept of hierarchy is defined in directed networks and look for rankings of nodes into classes, while the latter makes sense also for undirected networks and look for nested clusters of nodes. 

In this paper we consider the problem of the inference of hierarchies in directed networks via a class of metrics recently introduced and termed agony. Given a ranking of nodes into classes (i.e. an ordered partition), agony is a metric which penalises those links which are against the ranking, i.e. from a high rank to a low rank node. Different forms of penalisation lead to different types of agony. Once the agony function is chosen, one looks for the ranking of nodes which minimises it. Thus optimisation of agony is a non-parametric approach of hierarchy detection. Similarly to community detection with modularity, agony minimisation might be a challenging computational task, even if for some forms of the agony function exact or heuristic algorithms have been recently proposed (see the next Section for more details).

Here we focus our attention on the problem of resolution limiti when detecting ranking hierarchies with agony minimisation. Specifically, we ask when a given hierarchical structure in a network can be identified with agonies. To investigate the possible resolution limits we proceed in a way similar to what has been done for community detection \cite{bickel2009nonparametric,zhang2014scalable,newman2016community}. We introduce a class of random graphs, termed {\it Ranked Stochastic Block Models} (RSBM)\footnote{Following the comment above, we stress again that our RSBM are different from those recently proposed in \cite{peixoto2014hierarchical}, where the nested concept of hierarchy is adopted.} which is a specific subclass of Stochastic Block Models \cite{holland1983stochastic} with a tunable hierarchical structure and we study the resolution limit of hierarchy detection with agony in RSBM.

We find, both analytically and empirically, that agony succeeds in identifying hierarchies when the structure is strong. However we prove the existence of resolution thresholds in the model parameters such that beyond these thresholds agony minimisation identifies hierarchical structures which are different from the planted one. Using symmetry arguments we explore analytically alternative rankings, showing that they can have a smaller agony (higher hierarchy) than the planted one. These rankings are obtained by merging, splitting, or inverting the classes of the planted ranking. It is important to notice that, as we show numerically, the resolution limits are not due to the RSBM, but to the method. Once more, this is analogous to what observed in community detection with modularity maximisation. Finally we show that in some cases, iterating the optimisation on each class found in the first run of the agony minimisation, it is possible to improve significantly the recovery of the planted structure.

The paper is organised as follows. In \cref{sec:agony} we introduce the cost functions for agony, in \cref{sec:hsbm} we define the model for RSBM and we compute an estimate for the value of agonies of  graphs in the ensemble, in \cref{sec:optimalHier} we study the resolution limit for this class of graphs, and in \cref{sec:sim} we present some numerical simulations which support the analytical computations.
In \cref{sec:iter} we present some empirical examples on both real and synthetic networks to show that it is possible to partially overcome the resolution limit issues. Finally in \cref{sec:concl} we draw some conclusions.

\section{Agony}\label{sec:agony}

Let $G=(V,E)$ be a binary directed graph of $N\equiv |V|$ nodes and $m\equiv |E|$ links. A rank function  $r:V\to \{1,...,R\}$ associates each node to an integer number which indicates the position of the subset (or class) containing the node in the hierarchy. Thus a rank function generates an ordered partition\footnote{From this point, we will refer to the ordered partition induced by the rank function with the term ranking.} of the nodes into $R$ subsets $\mathcal{C}_i$ ($i\in \{1,...,R\}$) of size $n_i=|\mathcal{C}_i|$.

Once a ranking has been assigned to the graph $G$, a link between two nodes is classified as {\it forward} if it goes from a node in a class to one in a class with a strictly higher rank and {\it backward} otherwise. Identifying the optimal hierarchical structure in a directed graph means to find a ranking where the presence of 
 backward links is suitably penalised. The penalisation will in general depend on the number of backward links as well as on the distance in rank between the connected nodes. The penalisation is of course arbitrary and it is interesting to investigate the ability of different forms of penalisation in identifying hierarchies.

\begin{figure}[t]
\center
\subfigure[$A_1^*=0$, $h_1^*=1$]{\includegraphics[scale=0.42]{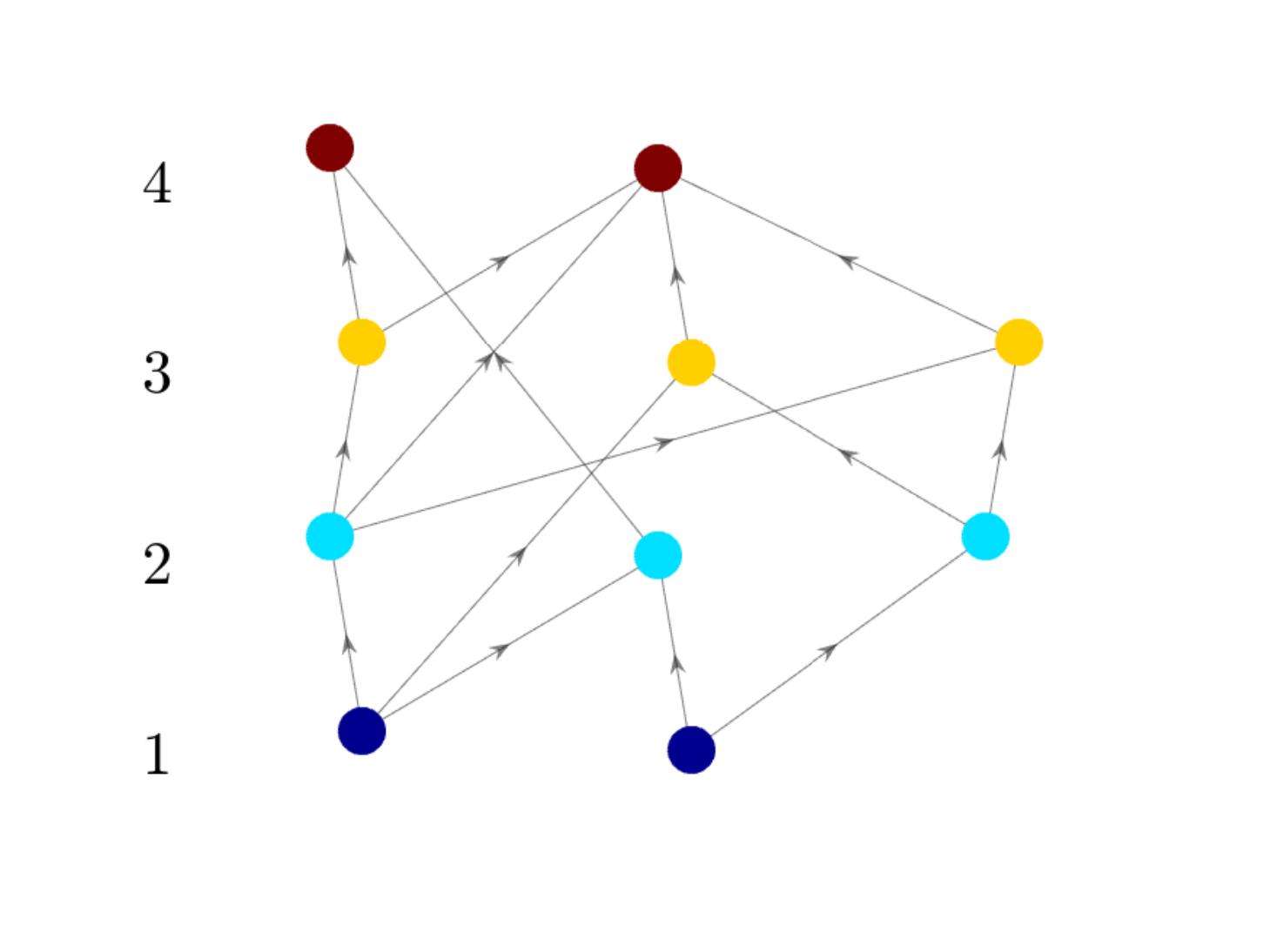}}
\subfigure[$A_1^*=11$, $h_1^*=0.5$]{\includegraphics[scale=0.42]{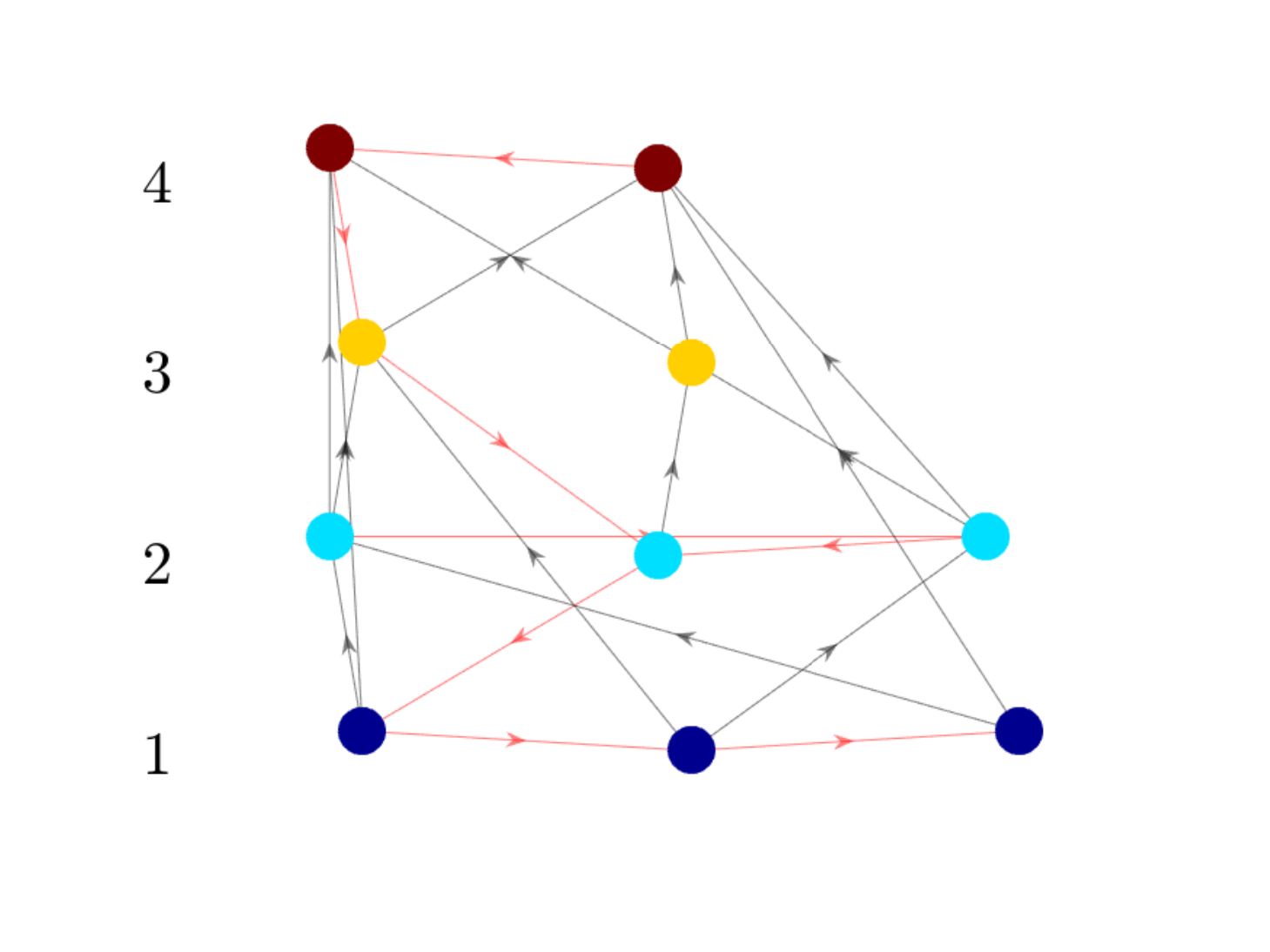}}
\caption{Optimal rank and agony ($d=1$) for simple graphs. The red links are the backward ones, those contributing to agony, and the black links are the forward ones. }
\label{fig:toy}
\end{figure}

The concept of agony in graphs was first introduced in \cite{Gupte2011} 
and it is the weighted cost of all the backward links in a ranking. More specifically, given a graph $G$ and a ranking $r$ the value of agony with respect to $r$ is given by:
\begin{equation}
A_f(G,r)=\sum_{(u,v)\in E} f(r(u)-r(v))\,,\label{eq:agony}
\end{equation}
where $f$ is a penalty function such that it is zero for negative argument and non decreasing otherwise.
We will consider here $f$ of the form
\[f_d(x)=\begin{cases}(x+1)^d& x\geq0\\
0&x<0\end{cases}\quad \quad\quad\quad d\geq 0\,,
\]
and we will denote the value of agony of the ranking $r$ on graph $G$ with $A_d(G,r)$.   
The agony of the graph is defined as the minimum value of agony with respect to all possible rankings on the nodes, i.e.
\begin{equation}
	A_d^*(G)=\min_{r\in \mathcal{R}}A_d(G,r)\,,\label{eq:astar}
\end{equation}
where $\mathcal{R}$ denotes the set of all rankings.
Figure \ref{fig:toy} shows two examples of optimal rankings for simple graphs and illustrates the difference between backward and forward links.

\begin{oss}
\begin{enumerate}
\item When the graph is a Directed Acyclical Graph (DAG), one can always find a ranking of the nodes such that there are no backward links (see \cite{Newman2014} for a simple routine to solve this problem), hence the value of agony of a DAG is 0, and we say the graph has a perfect hierarchy. 
\item Thanks to the minimisation, it is $0\leq A_d^*\leq m$, the upper bound being  the value of agony for the trivial ranking  where all the nodes are in the same class.
\item The exponent $d$ acts as a tuning parameter: when it increases, only rankings with stronger hierarchies are privileged over the trivial one.
\item The optimal ranking may be not unique, however there exists a routine to choose the ranking with the smallest number of classes among those with the optimal value of agony (see \cite{tatti2017tiers} for more details).
\end{enumerate}
\end{oss}
Finally, one can define the \emph{hierarchy} of a directed graph as
\begin{equation}
h_d^*(G)=1-\frac{A_d^*(G)}{m}\,.\label{eq:hstar}
\end{equation}
From the previous remark (ii) it follows that $0\leq h_d^*\leq 1$ where $h_d^*=1$ indicates a perfect hierarchy.


Once the penalisation has been chosen, the problem of finding the optimal ranking is quite complex. In its original version, agony was defined with the piecewise linear cost function, i.e $d=1$ in our notation. With this choice few exact algorithms to identify the optimal ranking of a graph are known \cite{Gupte2011,tatti2017tiers}. Ref. \cite{tatti2017tiers} considered the computational complexity of algorithms for generic $d$. The case $d=1$ is proven to be solved by an algorithm of polynomial complexity, while the case $d=0$ can be reformulated into the Feedback Arc Set problem (FAS)\cite{slater1961inconsistencies} which is known to be NP-hard, but for which some heuristics exist \cite{eades1993fast}. The intermediate cases, $0<d<1$, have concave cost functions, which also lead to a NP-hard problem according to \cite{tatti2017tiers}.
 The case $d>1$, instead, have a convex cost function which gives a problem of polynomial complexity. However, to the best of our knowledge, no algorithm is available at the moment for these latter cases.

One of the objectives of this paper is to investigate how the detected optimal ranking depends on the choice of the penalty function. For this reason we need to introduce a class of graphs which have a hierarchical structure and whose strength can be tuned by a suitable choice of parameters. This is what we do in the next Section. 

\section{Ranked Stochastic Block Model}\label{sec:hsbm}

Our ensemble of graphs belongs to the class of Stochastic Block Models (SBMs).
In this ensemble of graphs, nodes are partitioned into $R$ disjointed subsets and the probability of having a link between two nodes depends on the classes they belong to and it is independent of all the other pairs of nodes, i.e.
\[
\mathbb{P}[(u,v)\in E\,|\,u\in \mathcal{C}_i,\,v\in \mathcal{C}_j]=c_{ij}\,.
\]
The $R\times R$ matrix $C=\{c_{ij}\}_{ij}$ is called the {\it affinity matrix}. For our purpose we consider the directed version of SBMs, and $C$ is not symmetric. We choose a parametrisation of $C$ in order to keep the number of parameters small, which allows to have both analytical tractability, and  enough flexibility to model different types of hierarchies.

The ranking $r^{(p)}$, which we will refer to as \emph{planted ranking}, is defined so that it is consistent with the labelling in the affinity matrix, i.e.
\[
 r^{(p)}(\mathcal{C}_i)=i\quad i=1,\dots,R\,.\\
 \]
Note that, given the collection of subsets of nodes, any rank function with a range of values larger than $R-1$ would have a larger value of $A_d$.

Consider
\begin{align*}
p&=\mathbb{P}(\text{forward link towards a node in the nearest upper class})\,,\\
q&=\mathbb{P}(\text{forward link towards more distant classes})\,,\\
s&=\mathbb{P}(\text{backward link})
\,.
\end{align*}

This gives the affinity matrix
\[
C=\begin{bmatrix}
s  &  p     &            &            &        &\\
    &   \ddots  & \ddots &            &q\\
    &         &  \ddots  &  \ddots&        &\\
    &         &          & \ddots  & \ddots \\
    &      s   &            &           &  \ddots &   p  \\
    &         &            &           &          &    s 
\end{bmatrix}
\]

In order to have a true hierarchical structure we require that the parameters $p,q,s$ are such that
\begin{equation}\label{eq:hierarchy}
\mathbb{E}[\#\{\text{backward links}\}]\leq \mathbb{E}[\#\{\text{forward links}\}]\,.
\end{equation}
Define $\forall k\in 1,..,R$,
\[b_k=\sum_{i=1}^{R-k}n_in_{i+k}\,.
\]
For any pair $(i,j)$ ($i,j=1,....R$), the number of links between subset $i$ and $j$ $m_{i,j}$ follows a binomial distribution, $m_{i,j} \sim \text{Binom}(n_in_j, (C)_{i,j})$
, therefore the constraint \eqref{eq:hierarchy} is equivalent to
\[
s\sum_{k=0}^{R-1}b_k\leq pb_1+q\sum_{k=2}^{R-1}b_k\,.
\]
In the case of uniform cardinality of the subsets, $n_i=n \, \forall\,i$, which we will consider in the following, the inequality further simplifies to
\begin{equation}
s\leq s_{\max}:=\frac{2(R-1)}{R(R+1)}p+\frac{(R-2)(R-1)}{R(R+1)}q\label{eq:exBound}
\end{equation}
A SBM having the above structure and satisfying the constraint \eqref{eq:hierarchy} will be termed Ranked Stochastic Block Models RSBM$(p,q,s,R,\{n_i\})$. In the case of uniform cardinality, we denote briefly RSBM$(p,q,s,R)$. Since RSBMs are random graphs, different realisations of the model  give different values of agony and hierarchy. We will compute below the expected value of these quantities.

We estimate the expected value of $h_d(G,r^{(p)})$, the hierarchy of the planted ranking of RSBM graphs\footnote{We make a little abuse of notation indicating with $h_d$ the value $1-A_d/m$, i.e. we do not consider the minimization of agony. For this reason $h$ is not necessarily bounded between $0$ and $1$ as $h_d^*$.}.
Indicating with $\bar{h}_d^{(p)}$ the ensemble average of $h_d(G,r^{(p)})$, we obtain 
\begin{align*}
\bar{h}_d^{(p)}&=\mathbb{E}\left[1-\frac{1}{m}A_d(G,r^{(p)})\right]\,
=1-\sum_{i\geq j}(i-j+1)^de_{ij}\,,
\end{align*}
where 
$ e_{ij}=\mathbb{E}\left[\frac{m_{ij}}{m}\right]\,.$

In order to have closed form expressions we need to estimate the terms $e_{ij}$. 
We consider a second order Taylor expansion:
\begin{equation}
\mathbb{E}\left[\frac{m_{ij}}{m}\right]\approx \frac{\mathbb{E}[m_{ij}]}{\mathbb{E}[m]}-\frac{\cov(m_{ij},m)}{\mathbb{E}[m]^2}+\frac{\var(m)\mathbb{E}[m_{ij}]}{\mathbb{E}[m]^3}\,.\label{eq:tay2}
\end{equation}
If we assume that $n_i=O(N)\,\forall i$, then the last two terms in Eq. \eqref{eq:tay2} vanish when $N\to\infty$, hence
\[
e_{ij}\to  \frac{\mathbb{E}[m_{ij}]}{\mathbb{E}[m]}\quad\text{as}\quad N\to\infty\,.
\]
This gives the first order estimate for $\bar{h}_d^{(p)}$


\begin{align*}
\bar{h}_d^{(p)}&=1-\frac{\mathbb{E}[A_d(G,r^{(p)})]}{\mathbb{E}[m]}+o(N^{-1})\\
&=1-\frac{s\sum_{k=0}^{R-1}(k+1)^db_k}{pb_1+q\sum_{k=2}^{R-1}b_k+s\sum_{k=0}^{R-1}b_k}+o(N^{-1})\,.
\end{align*}
It is possible to compute higher order estimates or estimates based on exact expected values. The expressions are however less transparent and we find in simulations that first order estimates are quite accurate, thus in the following we use them.

\section{Looking for optimal hierarchies in RSBM}\label{sec:optimalHier}

RSBMs are constructed with a specific ranking, the planted one, which is determined by the choice of the classes and the model parameters. When minimising a generalised agony $A_d$ on realisations of such graphs, it is not {\it a priori} obvious that the optimal ranking is the planted one. We therefore ask the following question:

\medskip

{\it Given a RSBM$(p,q,s,R,\{n_i\})$, find the ranking $r$ which minimises the generalised agony $A_d$. In particular check when the planted ranking $r^{(p)}$ is optimal.}

\medskip

This is in general a complicated problem and we do not have a complete answer to this question, despite the fact that it is possible, at least for $d=1$, to find numerically the optimal ranking of a specific realisation of a RSBM. In order to simplify the problem, in this paper we will restrict our attention to the homogeneous case $n_i=N/R$, $\forall i$. Given the form of the affinity matrix and the homogeneity assumption, we expect that the optimal solution, when different from the planted one, preserves the homogeneity of the planted ranking. Possible boundary effects (for example the first and last class have different size from the other ones) are not considered and we expect to play a minor role when the number of planted classes is large. In any case in Section \ref{sec:sim} we use numerical simulations to test our intuition.


For this reason we shall compute the generalized agony of the following alternative rankings:
\begin{enumerate}
\item the number of classes changes either by merging adjacent classes or by splitting each class; due to homogeneity, merged or split classes have all the same size;
\item the rank is inverted, $r_j^{(i)}=r^{(p)}_{R-j+1}$, $\forall j=1,..,R$, i.e. nodes in highest ranks of the planted ranking are given lowest ranks in the alternative. Moreover we consider also the case when the number of classes is arbitrary, but again their size is assumed to be uniform.
\end{enumerate}
To distinguish between the two families of ranking, we will denote the former as \emph{direct}, in contrast with \emph{inverted} for the latter. 
For each of these alternative rankings we compute the value of $\bar h_d$ as a function of the number of classes and we look for the optimal one among these alternatives and the planted ranking. Clearly there is no guarantee that this will be the global optimum over all the possible rankings\footnote{To maintain this distinction, we will denote \emph{optimal} the ranking with highest value of $\bar h_d$ within the subset of alternatives just described, while we will always refer to the best  among all the rankings, i.e. that which gives $h_d^*$, as the \emph{global optimum}.}. We will see for example that numerical simulations of some RSBM indicate that the globally optimal ranking is a {\it partial} inversion of the planted hierarchy. However this analysis serves to show that planted ranking might not be globally optimal for some generalised agony and to provide an upper bound for the resolution threshold as well as getting intuition on the characteristics of the optimal ranking in a RSBM.

In the following we will focus on two regions of the parameter space of RSBMs:
\begin{itemize}
\item $p\geq q>s$, termed a twitter-like hierarchy;
\item $q=0,\,p\neq0$ termed a military-like hierarchy.
\end{itemize}
In the former hierarchy forward links can connect low rank nodes with nodes of any higher rank, while in the latter the forward links can connect a node only with nodes in a direct superior class. In both cases backward links can exists with a probability $s$. As we will see the global optimal ranking of the two hierarchies is quite different. 

Finally we consider the case
\[R=2^a,\,\tilde R=2^{a-b}\,,\]
where $\tilde R$ is the number of classes after splitting ($b<0$) or
merging ($b>0$). The parameters $a>1$ and $b<a$ are such that  $2^{a},\,2^{a-b}\in\mathbb{N}$. We denote the direct and inverted rankings with $2^{a-b}$ classes as $r^{(b)}$ and $r^{(i,b)}$ respectively.

We will focus our attention on the case $d=1$, $d=0$, and $d=2$. Results for other values of $d$ are left for a future paper.

\subsection{Agony with $d=1$}
In this case exact algorithms for its optimisation are known, allowing the comparison of calculations with numerical simulations.

Provided that the constraints in \eqref{eq:exBound} are satisfied, one can easily verify that $\forall\,b<0$
\begin{align*}
\mathbb{E}[A_1(G,r^{(b)})]&>\mathbb{E}[A_1(G,r^{(p)})]\,\\
\mathbb{E}[A_1(G,r^{(i,b)})]&>\mathbb{E}[A_1(G,r^{(p)})]\,
\end{align*}
i.e. splitting is never optimal, neither in the direct nor in the inverted ranking.

As for merging $(b>0)$, the first order estimate of $\bar h_1$ is given by
\begin{equation}
\bar{h}_{1}{(b;\,p,q,s,a)}=\frac{2^{-b}(2^a-2^b)(6p+3(-2+2^{a+b})q-2^a(2^a+2^b)s)}{3 \left(2^a (2 p-3 q+s)+4^a (q+s)-2 p+2 q\right)}\,.\label{eq:h-d1}
\end{equation}
Similarly, one can write the estimate for the value of hierarchy of the inverted ranking
\[\bar{h}^{(i)}_{1}{(b;\,p,q,s,a)}=\frac{2^{-b} \left(2^b-2^a\right) \left(2^{a+b} (q-3 s)+\left(4^a-6\right) q+6 p\right)}{3 \left(2^a (2 p-3 q+s)+4^a (q+s)-2 p+2 q\right)}\,.
\]
In this notation $p,\,q,\,s,\,a$ are the parameters of the RSBM, while $b$ refers to the modified ranking $r^{(b)}$ or $r^{(i,b)}$. Moreover it is clearly $\bar{h}_1(b=0)=\bar h_1^{(p)}$.

In the twitter-like hierarchy ($p\geq q>s$) it is $\bar{h}_1^{(p)}>\bar{h}^{(i)}_{1}{(b;\,p,q,s,a)}$, i.e. the inverted ranking is never optimal. Merging, instead, can give rankings with higher hierarchy than the planted ranking. 

\begin{figure}[t]
\center
\subfigure[]{\includegraphics[scale=0.4]{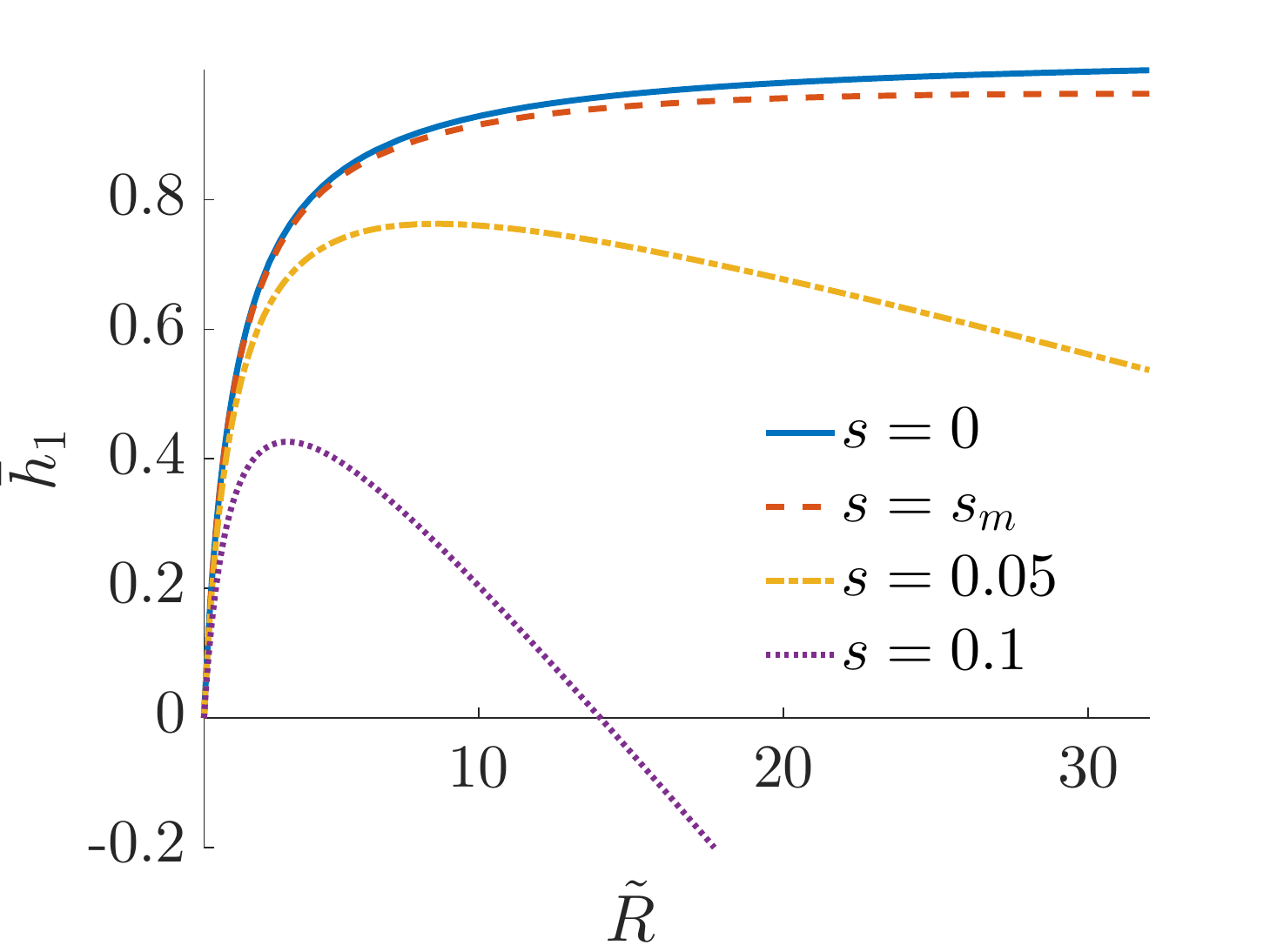}}
\subfigure[]{\includegraphics[scale=0.4]{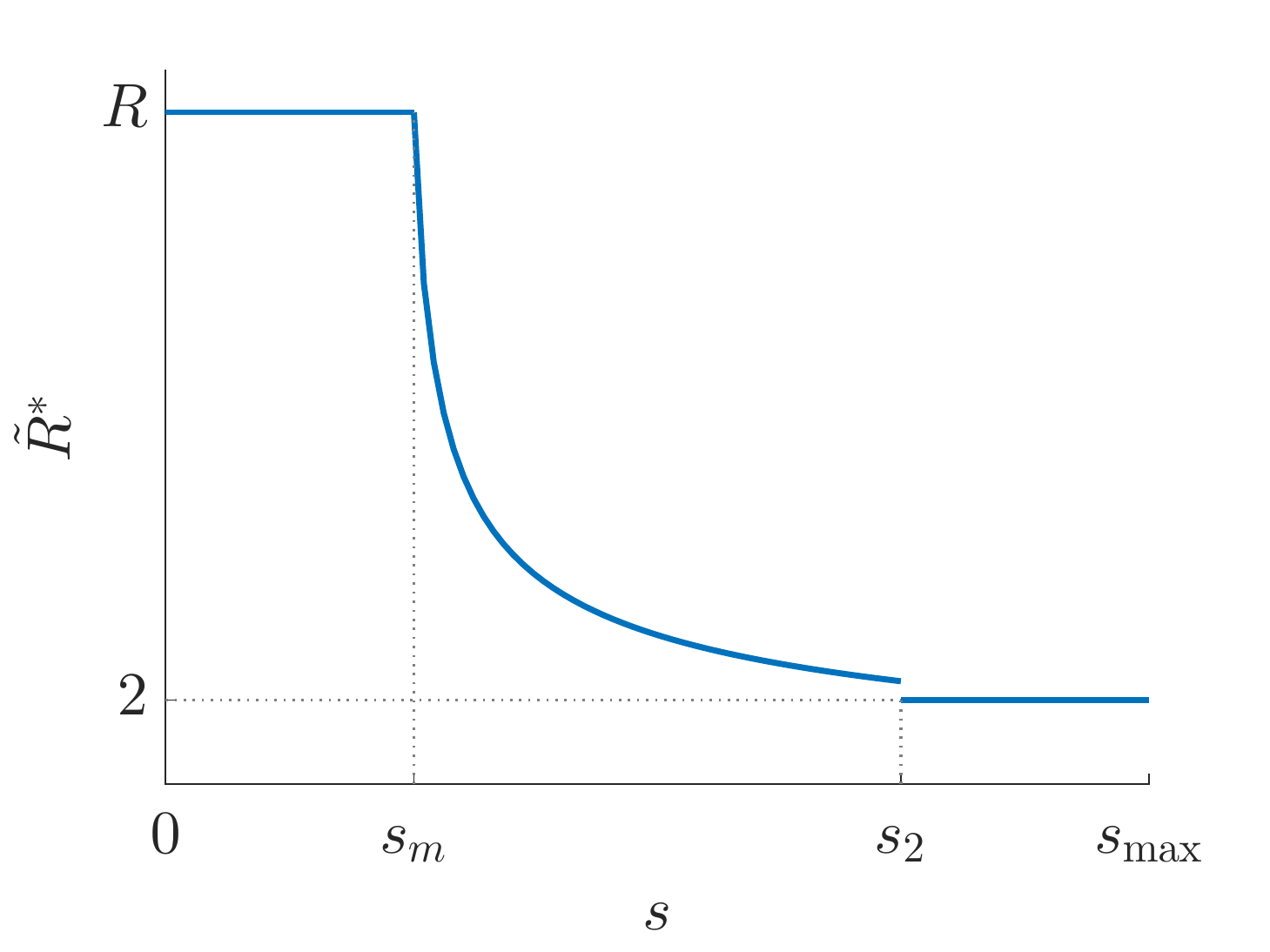}}
\caption{Panel (a)  shows the value of the estimate of $h_1$ for different values of $s$ as a function of the number of classes, $\tilde R$, for twitter-like graphs with parameters $p=q=0.5,\,R=32$. Panel (b) gives a schematic representation of the estimated optimal number of classes $\tilde R^*$ as $s$ varies.}
\label{fig:hb}
\end{figure}

To show this, in the left panel of Figure \ref{fig:hb} we plot the behaviour of $\bar{h}_1(b)$ as a function of the number of classes\footnote{We plot the variable $\tilde{R}$ as a continuous variable to help the interpretation of the observed behaviour.}, $\tilde{R}=2^{a-b}$, after merging. Each line is associated to a $RSBM(p,q,s,R)$. The parameters  $p=q=0.5$, $R=32$ are fixed, while different curves refer to different values of $s$. When $s$ is small the maximum value of $\bar{h}_1$ is correctly identified at $\tilde R=R$. Above a critical value $s_m$ of the parameter describing the probability of a backward link, the planted ranking is no longer optimal and merging classes gives a ranking with higher hierarchy. Notice that for $s>s_{\bar h_1^{(p)}=0}$, the hierarchy $\bar h_1^{(p)}$ of the planted ranking becomes negative. This might seem counterintuitive since we showed before that $h^*\in [0,1]$. The condition $\bar{h}_1^{(p)}<0$ simply means that putting all the nodes in the same class has a higher hierarchy than the one of the planted ranking when $s>s_{\bar h^{(p)}=0}$.

The right panel of Figure \ref{fig:hb} shows the optimal number of classes $\tilde R^*$ as a function of $s$. As explained, when $s<s_m$ it is $\tilde R^*=R$, while after this value the optimal number of classes decreases and in the limit $s=s_{\max}$ it is $\tilde R^*=2$. Therefore the value $s_m$ sets a {\it resolution threshold}, since twitter-like graphs with a probability of backward links larger than $s_m$ will not be correctly identified by agony with $d=1$. More precisely $s_m$ is an upper bound of the resolution threshold, since other rankings, not considered here, could have higher hierarchy than the planted and the merged ones when $s<s_m$. 

Interestingly for large number of classes $R$ the resolution threshold scales as $s_m\sim(6p-3q)/R^2$, i.e. the more communities are present the more it is difficult to detect them. The same happens for large networks ($N\to +\infty$). Taking the number of classes constant and letting $p$ and $q$ scale as $1/N$ to keep the connectivity fixed, one immediately sees that $s_m=O(N^{-1})$, i.e. for large networks and fixed number of classes the detectable structures are those with very strong hierarchical structure. 
Thus agony with $d=1$ has strong resolution limits for large graphs, similarly to what happens with modularity and community detection.


\begin{figure}[t]
\center
\subfigure[]{\includegraphics[scale=0.4]{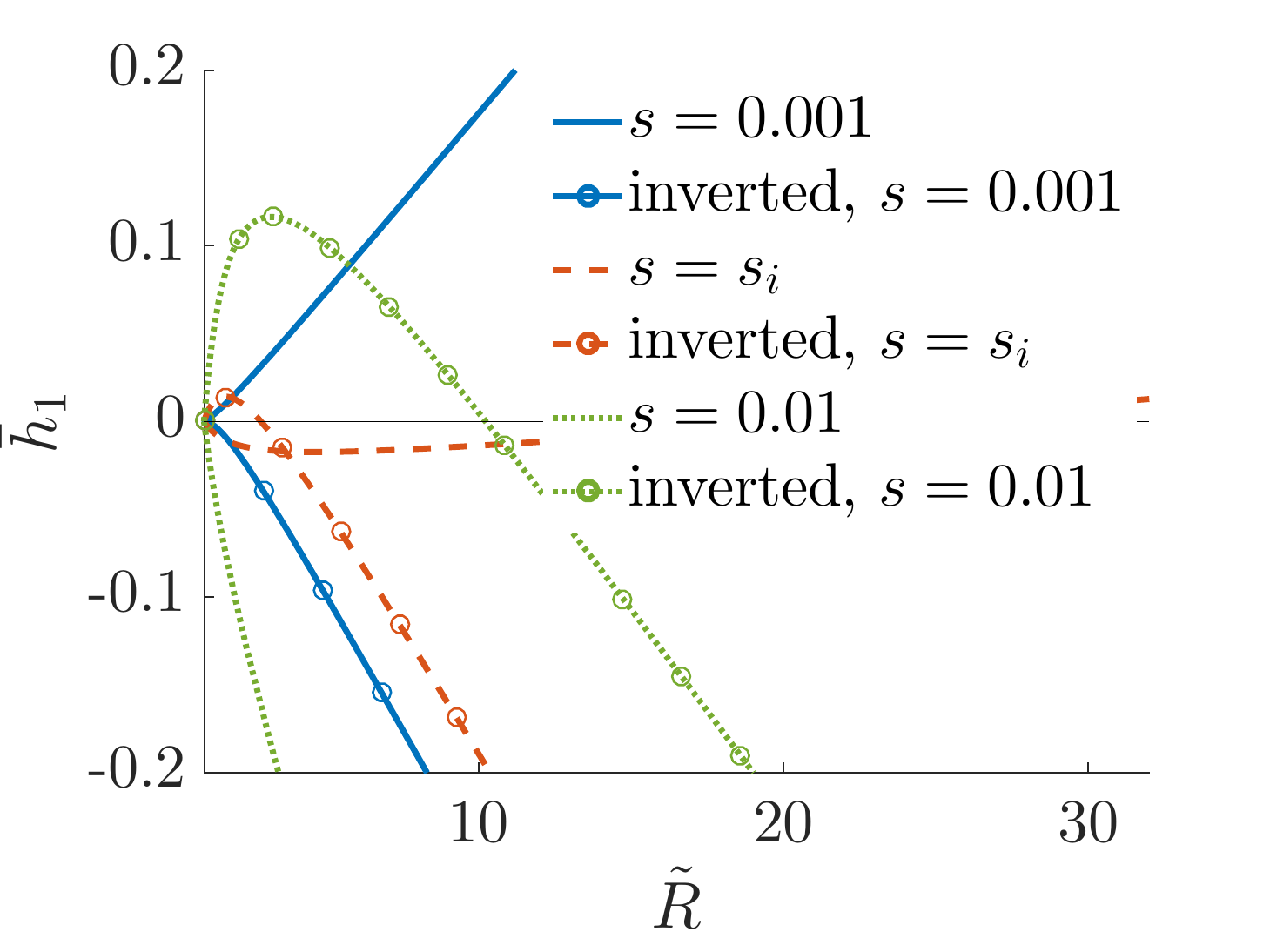}}
\subfigure[]{\includegraphics[scale=0.4]{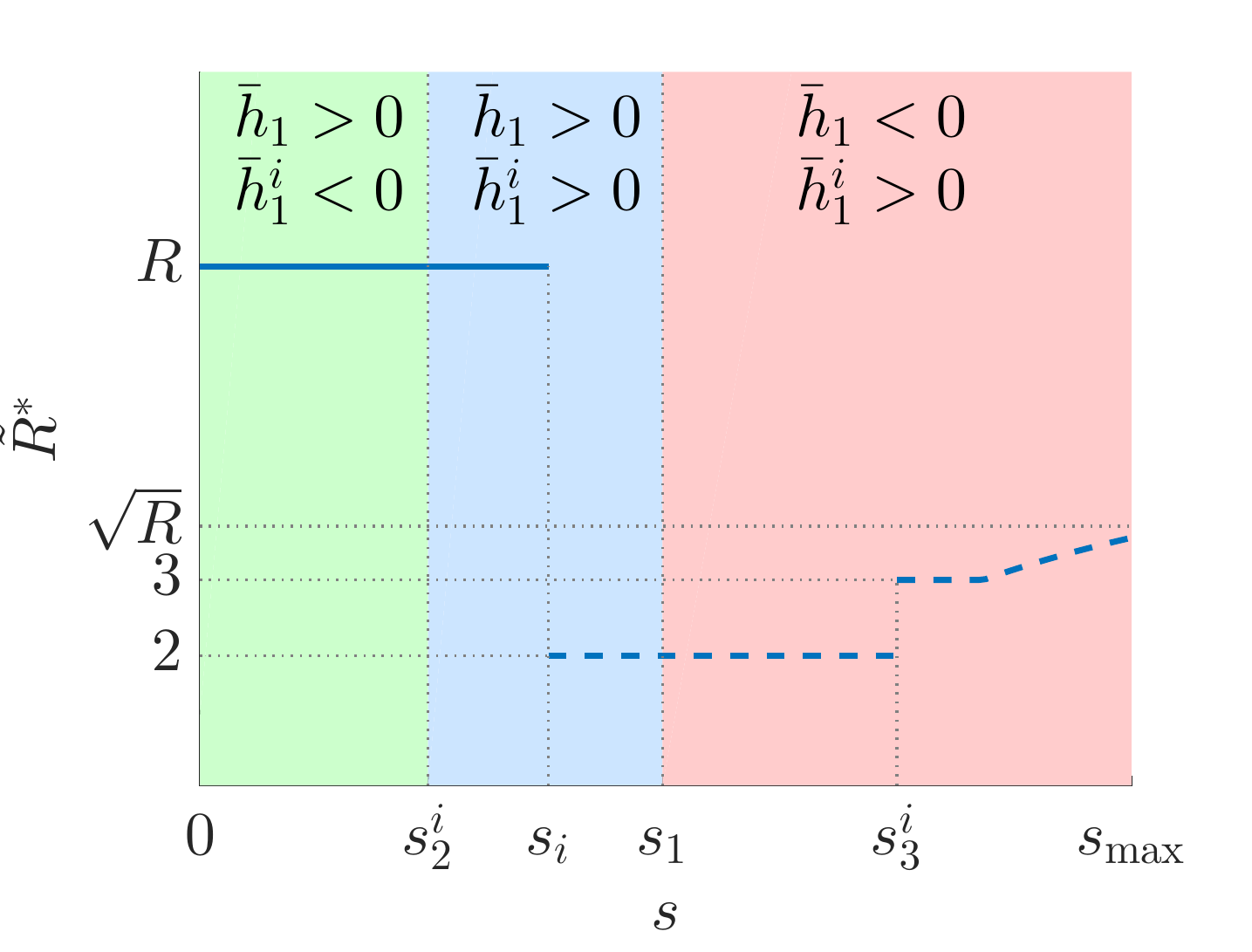}}
\caption{Panel (a) shows how depending on the value of $s$ the inverted rank can give a higher value of $\bar h$ than the planted rank in military-like graph with parameters $p=0.5,\,q=0,\,R=32$. Panel (b) gives a schematic representation of the estimated optimal number of classes $\tilde R^*$ as $s$ varies, dashed lines are associated to the inverted rank.}
\label{fig:hbq0}
\end{figure}

The situation is more complex in the military-like hierarchy ($q=0$) because for large $s$ inverted rankings become better than direct ones.  To show this, we refer to the left panel of Figure \ref{fig:hbq0}, which is the analogous of left panel of Figure \ref{fig:hb}. In this case, alongside $\bar h_1(b)$ we also plot $\bar h^{(i)}_1(b)$, with matching line colours to distinguish those associated to the same values of $s$, and circles to identify $\bar h^{(i)}_1$. In all cases we chose $p=0.5$ and $R=32$. For small values of $s$ (solid blue lines), $\bar h_1$ is convex in $\tilde R$ and has its maximum at $\tilde{R}=R$, whereas $\bar h^{(i)}_1$ is negative for inverted rankings different from the trivial one. Thus in this regime the planted ranking is optimal. 
When $s$ reaches the critical value $s_i$ (dashed red lines), the optimal choices for both the direct and inverted rankings give the same value of hierarchy.
For higher $s$ (dotted green lines) the only direct ranking with non negative hierarchy is the trivial one, i.e $\tilde{R}=1$, while the inverted rankings are (strictly) positive for a suitable choice of $b$. Therefore in this regime inverted rankings outperfom the planted one.

The right panel of Figure \ref{fig:hbq0} shows the optimal number of classes $\tilde R^*$ as a function of $s$ together with an indication of the sign of the hierarchy of the optimal direct and inverted ranking. For $s<s_2^i$ the hierarchy of the optimal direct ranking is positive and the one of the optimal negative ranking is negative, for $s_2^i<s<s_1$ they are both positive, while for $s_1<s<s_{\max}$ the inverted optimal hierarchy is positive and the optimal direct one is negative. Thus for $s<s_i$ the optimal ranking is direct and coincides with the planted one, while after this value the inverted ranking with two classes becomes optimal. This is true in the region $s_i<s<s_3^i$ after which the inverted  ranking with three classes becomes optimal. By increasing $s$ further, the optimal ranking is always inverted with an increasing number of classes up to a value smaller or equal to $\sqrt{R}$ for $s=s_{\max}$. Therefore for the military-like hierarchy the resolution threshold is $s_i$ which for large $R$ scales as $6p/R^2$, displaying a resolution limit similar to the twitter-like hierarchy, both for large number of classes $R$ and for large graphs ($N\to \infty$).



\medskip

We summarise the results for $d=1$ in the following proposition. 

\begin{proposition}\label{prop:d1}
When $d=1$ and  $p\geq q>s$, \textcolor{blue}{(Twitter hierarchy)} the first order estimate for the optimal value of $h$
\begin{equation}
\bar{h}_1^*=\begin{cases}
 \bar{h}_1^{(p)}& s\leq s_m
   \\
 \bar{h}_1{(b=b^*)}  & s_m<s<s_2 \\
 \bar{h}_1{(b=2)}  & s\geq s_2\,,
\end{cases}
\label{eq:hstar-d1}
\end{equation}
where
\begin{align*}
s_m&=\frac{6 \left(2^a-1\right) p-3 \left(2^a-2\right) q}{2^a-4^a+8^a}\,,\quad
s_2=\frac{3}{7}\frac{\left(4^a-12\right) q+12 p}{ 4^{a}}\,,\,
\\b^*&=\frac{1}{2}\log_2{\frac{2^{2a}s+6(q-p)}{3q-s}}\,.
\end{align*}
Furthermore, when $q=0$,  \textcolor{blue}{(Military hierarchy)}
\[
\bar{h}_1^*
=\begin{cases}
\bar{h}_1^{(p)}
& s\leq  s_i \\
\bar{h}_1^{(i)}{(b=a-1;\, q=0)} & s_i<s_3^i \\
\bar{h}_1^{(i)}{(b=b^{i,*};\,q=0)} &  s>s_3^i\,,
\end{cases}
\]
where
\begin{align*}
s_3^i=\frac{12}{2^{2a}}p\,,\quad
s_i= \frac{12 p}{3\ 2^a+2^{2 a+1}-2}\,,\quad
b^{i,*}=\frac{1}{2}log_2{\frac{2p}{s}}\,.
\end{align*}
\end{proposition}
The proof and the extended expression for $\bar{h}^*_1$ are given in Appendix \ref{ap}. 

In conclusion, we explicitly showed that for RSBMs there exist alternative rankings with a smaller agony ($d=1$) than the planted one. The merging of the classes for the twitter hierarchy is due to fact that for a large number of classes it might be more convenient to aggregate classes paying a penalty equal to one than to leave them separate but paying a higher penalty for the distant backward links. Similarly, for the military hierarchy, when the number of backward links is relatively large, it is more convenient (in terms of agony) to invert the ranking because forward links do not enter the cost minimization. Thus even if $p$ is much larger than $s$ and the number of forward links is much larger than the number of the backward links, it is more convenient to invert the ranking to avoid to pay large penalties of backward links between very distant classes. 

Thus our results depend on the choice of the penalization function and on the choice of the affinity matrix. In the next Subsection we show indeed that a very different result is obtained for $d=0$. Changing the affinity matrix, for example introducing a probability of backward links which depends on the distance between classes, and changing the penalty function by including the negative cost of forward links is left for a future study.

\subsection{Agony with $d=0$}
This case corresponds to the FAS problem. The optimal ranking is obtained when each node is in a different class, $\tilde R=N$, and the inverted ranking is never optimal as stated by the following:
\begin{proposition}\label{p:d0}
When $d=0$, $\forall\,RSBM(p,q,s,R=2^a)$ the optimal value for the first order estimate of $h$ is given by (for both  \textcolor{blue}{Twitter and Military hierarchy})  
\[\bar{h}_0^*=\bar{h}_0\left(b=-\log_2{\frac{N}{R}}\right)\geq \frac{1}{2}\,.\]
\end{proposition}
See Appendix \ref{ap} for the proof.
The reason for this result is that backward links are weighted in the same way irrespectively from the distance between the ranks of the nodes connected by the link. Thus, for example, the naive ranking with all nodes in one class has a agony equal to the number of links, while the ranking where each node is in one class has an agony equal to the number of backward links, which is smaller than the total number of links. 

Finally we note that the value of $\bar{h}_0$ increases very slowly when $\tilde{R}$ approaches $N$, so in specific realisations of the RSBM the optimal ranking can have a number of classes smaller than $N$.

\subsection{Agony with $d=2$}
Finally, we consider the case of $d=2$. Similarly to the case $d=1$, splitting is never optimal, both for the direct and inverted rankings, while merging can give rankings with higher value of $\bar h_2$ than the planted one. One can proceed as before, considering the expressions for the alternative rankings when $b>0$:
{\small\[
\bar{h}_2(b;\,p,q,s,a)=
-\frac{2^{-2 b-1} \left(2^a-2^b\right) \left(2^{a+2 b+1} (2 s-3 q)+ 5 s 2^{2 a+b}+8^a s-3\ 2^{b+2} (p-q)\right)}{3 \left(2^a (2 p-3 q+s)+4^a (q+s)-2 p+2 q\right)}\,,
\]
}
and
{\small\[
\bar{h}_2^{(i)}(b;\,p,q,s,a)=
\frac{2^{-2 b-1} \left(2^b-2^a\right) \left(2^{a+2 b+1} (2 q-3 s)+\left(5\ 4^a-36\right) 2^b q+8^a q+9\ 2^{b+2} p\right)}{3 \left(2^a (2 p-3 q+s)+4^a (q+s)-2 p+2 q\right)}\,.
\]
}
As before we describe the behaviour for the two considered hierarchies and then we state the proposition summarizing our results. 
For the twitter-like hierarchy ($p\geq q>s$), the behavior is similar to the $d=1$ case. Since $\bar h_2^{(p)}> \bar h_2^{(i)}(b)$, $\forall b$, inverted rankings are never optimal. The planted ranking is optimal up to the critical value $s_{2,m}$ for the probability of backward links. After that, merged rankings outperform the planted one, and the number of classes decreases with $s$. When $s_{2,1}<s\leq s_{\max}$ the optimal choice is the trivial ranking, i.e. $\tilde{R}=1,h_2=0$. Despite the similarity with the $d=1$ case, the resolution threshold is now higher, since it can be shown that $s_{2,m}\le s_m$. Moreover, while, as noted before, in the $d=1$ case $s_{m}=O(R^{-2})$, in the $d=2$ case the resolution threshold is not only stricter but also it decreases faster as the number of classes increases, since it scales as $s_{2,m}\sim \frac{2p-q}{2R^3}=O(R^{-3})$. Finally, when $d=2$ the large $s$ case has the trivial ranking as the optimal one, whereas in the $d=1$ case the optimal ranking has two classes.

For the military-like hierarchy ($q=0$), the planted ranking is proven to be optimal with respect to the direct rankings up to the critical value $s_{2,1}^0$. After this value the optimal choice is the trivial ranking. Then 
when $s>s_{2,2}^i$ it becomes optimal to merge inverted rankings and the optimal number of classes increases with $s$, starting from $\tilde R=2$. Differently from the case $d=1$, in this case it holds $s_{2,1}^0<s_{2,2}^i$, hence for $s\in (s_{2,1}^0,\,s_{2,2}^i)$ the optimal rank is the trivial one, and the resolution threshold is given by $s_{2,1}^0$, which scales as $12 p/R^3$, while  inverted rankings are to be preferred for any $s>s_{2,2}^i$.

We summarise the results for $d=2$ in the following proposition.

\begin{proposition}\label{p:d2}
When $d=2$ and  $p\geq q >s$ \textcolor{blue}{(Twitter hierarchy)}, the first order estimate for the optimal value of $h$
\[\bar{h}_2^*=\begin{cases}
\bar{h}_2^{(p)}& s\leq s_{2,m}\\
\bar{h}_2(b=b_2^*)& s_{2,m}\leq s\leq s_{2,1}\\
0 & s> s_{2,1}\,,
\end{cases}
\]
where
\begin{align*}
s_{2,m}=\frac{6 \left(2^{1-a} (q-p)+2 p-q\right)}{-3\ 2^a+2^{3 a+1}+4^a+4}\,,\quad
s_{2,1}=\frac{2^{2 a} q+4 p-4 q}{3\,2^{2 a}}\,
\end{align*}
and $b_2^*$ is given in Eq. \eqref{eq:b2star} in Appendix \ref{ap}.

Furthermore, when $q=0$ \textcolor{blue}{(Military hierarchy)},
\[
\bar{h}_2^{*}=\begin{cases}
\bar{h}_2^{(p)}& s<s_{2,1}^0\\
0& s_{2,1}^0\leq s\leq s_{2,2}^i\\
\bar{h}_2^{(i)}(b=a-1;q=0)&s_{2,2}^i<s<s_{2,3}^i\\
\bar{h}_2^{(i)}(b=b_2^{i,*};q=0)&s\geq s_{2,3}^i\,,
\end{cases}
\]
where
\begin{align*}
s_{2,1}^0=\frac{3\ 2^{2} p}{2^a(5\ 2^a+4^a+4)}\,,\,
s_{2,2}^i=\frac{12}{2^{2a}} p\,,\,
s_{2,3}^i=3 s_{2,2}^i\,,\,
b_2^{i,*}=\frac{1}{2}\log_2 \left(\frac{6 p}{s}\right)\,.
\end{align*}
\end{proposition}

With this last proposition we showed that hierarchy detection with quadratic cost function has a behaviour very similar to the linear case
. However the resolution limits we highlighted before escalates in this case, and, as a result, only very strong hierarchies are detected correctly when the number of class is large. The same computations can be done also for greater integers $d$, for which the sums in the estimates of agony have a closed formula. Intuitively as $d\in \mathbb N$ increases, backward links to distant classes are given a larger penalisation, hence rankings with merged classes become more convenient than the planted one even for smaller values of $s$. In other words agonies with $d>1$ are strongly suboptimal and are able to identify very strong structures.  

Following this remark, better candidates as penalty functions are likely those with $0<d<1$. For at least some of those $d$ one can expect to soften the resolution limits associated to integer $d$. However the approach to study the regime cannot rely on analytical formulae.

\section{Numerical Simulations}\label{sec:sim}
In this Section we show the results of numerical simulations to test the  propositions we presented before. This is important for two reasons. First, to show that the guessed rankings, obtained by merging, splitting, or inverting the planted one, are indeed the optimal ones or have a hierarchy close to the optimal one. Second, to prove that the first order approximation and other simplifying assumptions give analytic expressions close to numerical simulations.

We use \emph{igraph} \cite{igraph} to sample a graph from the RSBM ensemble. For computing agony we use the algorithm described in \cite{tatti2017tiers}, which we will refer to it as {\it agony} (in italics) for brevity, and which gives the exact solution for the optimisation problem when $d=1$. Finally, we use the MCMC algorithm in the \emph{GraphTool} \cite{peixoto_graph-tool_2014} package for the inference of the SBM (without constraint on the structure of the affinity matrix). 

We perform the same analysis with different choices for the parameters $p,q,s,R,N$ and the results are consistent, thus in the following we present only representative cases. We use the adjusted Rand Index (RI) \cite{Hubert1985} to measure the similarity between the planted and the inferred ranking. The RI is $0$ between independent rankings and $1$ when each pair of elements that are in the same class in one ranking are also in the same class in the other\footnote{Ordering of classes does not matter in computing RI, thus the RI between a ranking and its inverted version is 1. Nevertheless we checked that high values of RI do not correspond to inverted rankings.}.

\subsection{Twitter-like hierarchy}

\begin{figure}[htp]
\center
\includegraphics[scale=0.5]{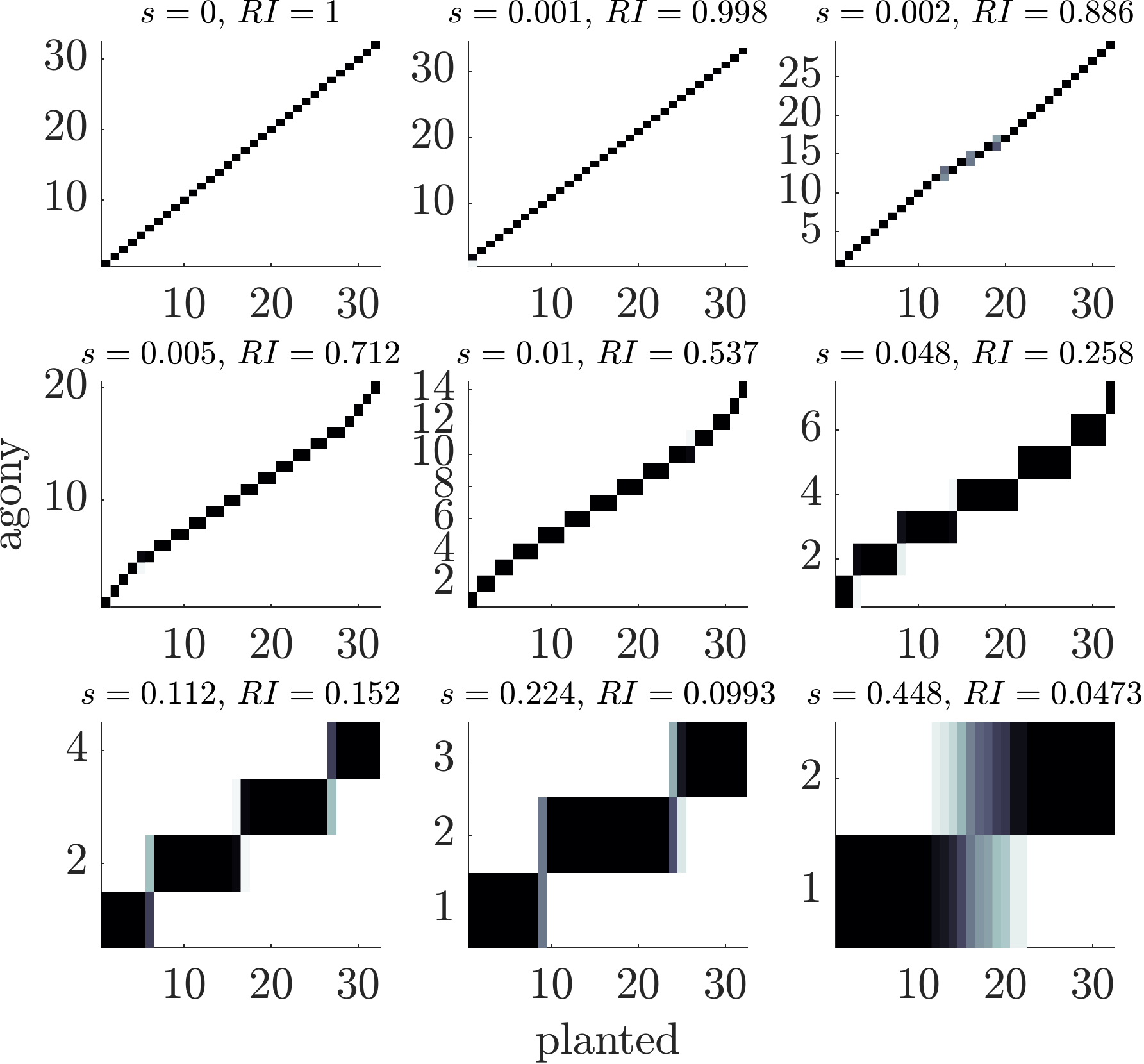}
\caption{Heat maps comparing the planted ranking with the ranking inferred with {\it agony} for twitter-like hierarchy. In each panel a square in position $(i,j)$ contains the number of nodes that belong to class $i$ in the planted rank and are placed in class $j$ by {\it agony}: the darker the colour, the higher the number. The parameters are $p=q=0.5,\,R=32$ and $9$ values of $s$. Each plot refers to a single realisation from the ensemble.}
\label{fig:hm}
\end{figure}

We generate RSBM with parameters $p=0.5,\,q=0.5,\,R=32 ,\, \frac{N}{R}=128,\,N=2^{12}=4096$, and we vary the value of $s$. 
\cref{fig:hm} shows the heat maps of the classes found by {\it agony} for different values of $s$. The heat-maps are constructed as follow: a square in position $(i,j)$ refers to the number of nodes that belong to class $i$ in the planted rank and are placed in class $j$ by {\it agony}: the darker the colour, the higher the number. For small $s$ (almost DAG structures) the algorithm recovers faithfully the planted ranking and the RI is high. When the hierarchical structure becomes weaker, the ranking obtained by agony is the merging of contiguous classes in the hierarchy, as postulated in the theoretical part above.  For this choice of $p,q,R$ the resolution threshold for $s$ is $s_m=0.00151$ consistently with our simulations. As we predicted, classes merge more and more when $s$ increases. The inferred rankings are close to uniform, and the main exception is the first and last class which are smaller than the other ones.

\begin{figure}[t]
\center
\includegraphics[scale=0.5]{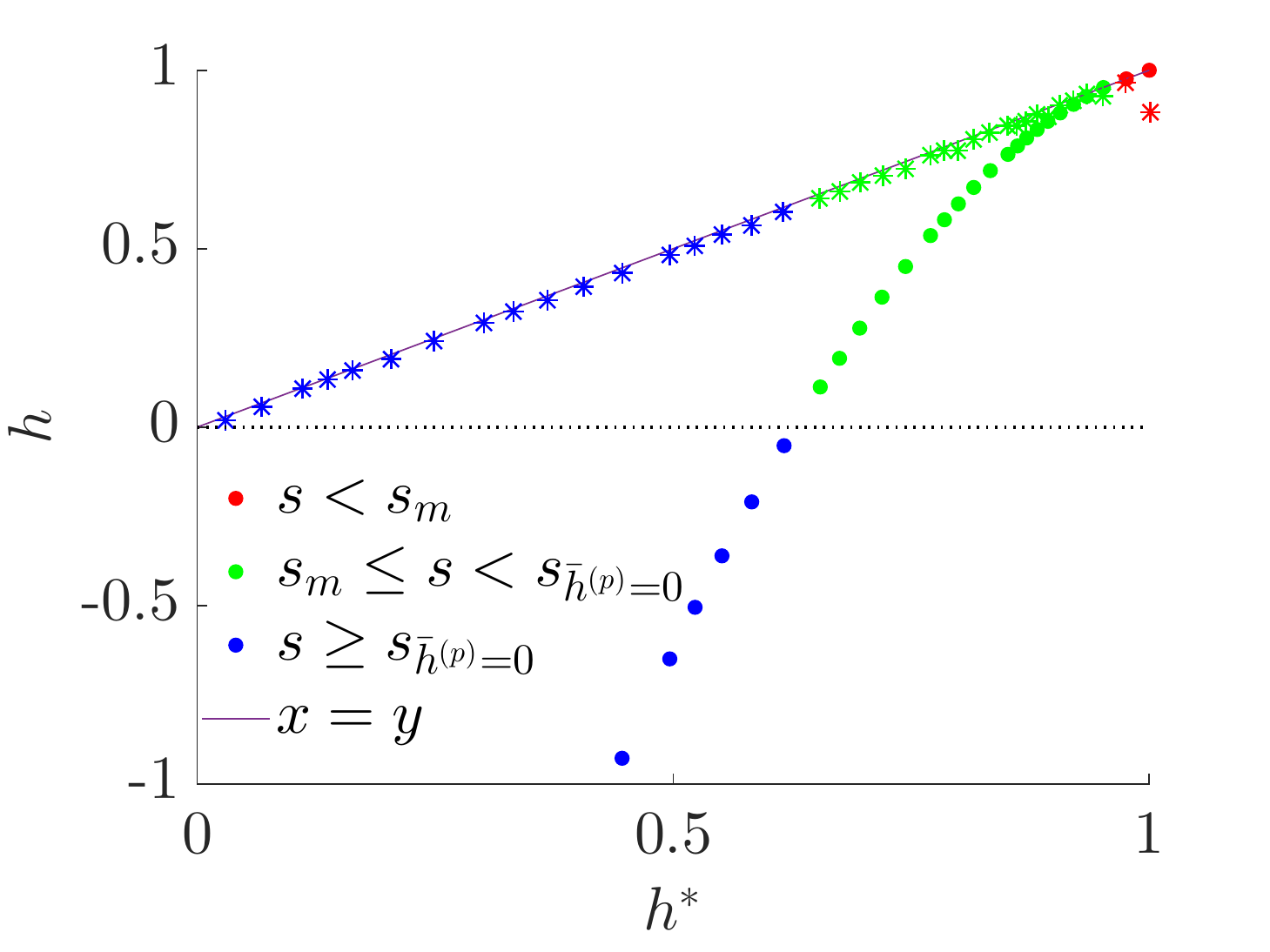}
\caption{Comparison of hierarchies for twitter-like HSBMs. The parameters are $p=q=0.5,\,R=32$, $s$ varies in $[0,s_{\max}]$, with $s_{\max}=0.448$. Each point refers to a single realisation of the ensemble. The circles represent the pairs $(h_1^*,h_1^{(p)})$, i.e. the optimal hierarchy $h_1^*$ computed with {\it agony} and the one of the planted hierarchy $h_1^{(p)}$. The stars represent $(h_1^*,\bar{h}_1^*)$ where $\bar{h}_1^*$ is the theoretical hierarchy of Eq. \eqref{eq:hstar-d1} with the parameters of the SBM estimated via \emph{GraphTool}. Finally, $s_m$ is the theoretical resolution threshold and $s_{\bar h^{(p)}=0}$ is the theoretical value of $s$ for which the estimate for the planted hierarchy is zero.}
\label{fig:hh}
\end{figure}

We show numerically that the ranking we proposed as optimal in the previous Section has indeed a value of hierarchy very close to the one obtained from simulations. In \cref{fig:hh} we show a scatter plot of the true value of $h_1^*$ computed with \emph{agony} on the simulated graphs  against the hierarchy of the planted rank $h_1^{(p)}$ (circles), and against $\bar{h}^*_1$, the hierarchy computed with Eq. \eqref{eq:hstar-d1}(stars). To evaluate the latter we use the coefficients of the RSBM estimated from the sample graph with \emph{GraphTool}. We estimate $p=q$ and $s$ as the average elements of the inferred affinity matrix on the corresponding classes and we leave free the number of classes. 
For $s<s_m$ (red symbols) the two methods agree and give a value of hierarchy consistent with the real one.  When $s>s_m$ (green and blue symbols depending on whether $s$ is smaller or larger of $s_{\bar h^{(p)}=0}$) the hierarchy of the planted ranking is significantly smaller than $h^*_1$, showing that another ranking is optimal. This has a value of hierarchy which is very close to the one computed from Eq. \eqref{eq:hstar-d1}, even when the coefficients of the RSBM are estimated from data\footnote{It is interesting to note that this is true also for $s$ very close to $s_{max}$ where the number of classes detected by \emph{GraphTool} is significantly smaller than $R$. This is due to the fact that the analytical expression in Eq. \cref{eq:hstar-d1} of the value of hierarchy  of the merged ranking depends weakly on the number of classes.}. This is a strong indication that the ranking we suggested, and obtained by merging the classes, has a value of hierarchy which is indeed very close to the globally optimal one.
In conclusion, the planted hierarchy is optimal for a very small range of values of $s$ and, as we expected, it gives negative values of $h_1$ when $s$ is large enough. On the other side, our estimate for optimal $h_1$ is accurate for all the value of $s$ considered.

\begin{figure}[htp]
\center
\includegraphics[scale=0.6]{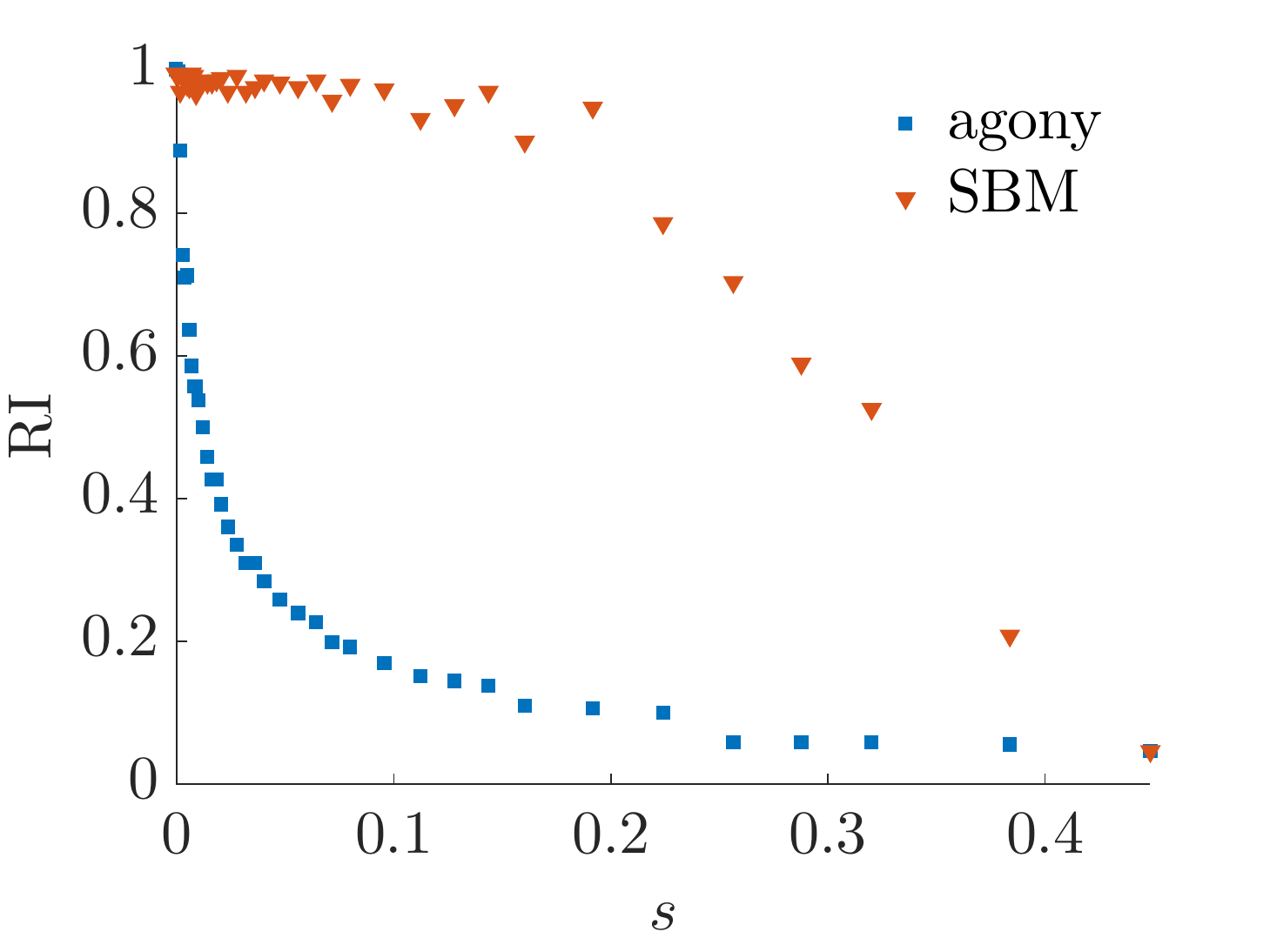}
\caption{The figure shows the value of the Rand Index between the planted ranking and the inferred ones. The blue squares considers the ranking obtained  with {\it agony} (hence $d=1$), while the red triangles considers the ranking obtained with a RSBM fit via {\it GraphTool}. The parameters of the twitter-like hierarchy are $p=q=0.5,\,R=32$, $s$ varies in $[0,s_{\max}]$, with $s_{\max}=0.448$, and each point refers to a single realisation of the ensemble.}
\label{fig:ri}
\end{figure}

Finally in \cref{fig:ri} we show that the resolution problem is due to the choice of the method (agony with $d=1$) and not necessarily to the model itself. In fact it is well known that SBM have a resolution threshold both when inference is done using Maximum Likelihood methods \cite{decelle2011asymptotic} and spectral methods \cite{Nadakuditi2012}. To this end we infer a SBM on the adjacency matrix, keeping free the number of classes (see \cite{peixoto_graph-tool_2014} for the model selection adopted by \emph{GraphTool}) and we compute the RI of the planted ranking versus the one obtained with {\it agony} and the SBM fit. The result is shown in \cref{fig:ri} for different values of $s$. We see that the SBM fit outperforms {\it agony}. It is clear that, since we are using SBM for generating the graphs, its fitting will be better. However what we want to stress is that there is remarkably wide interval of values of $s$ for which \emph{agony} is not able to detect a hierarchical structure even if it is strong enough to be detected by another method. Hence the limit in resolution is not embedded in the RSBM but in the objective function associated to agony. 

\subsection{Military-like hierarchy}

\begin{figure}[htp]
\center
\includegraphics[scale=0.6]{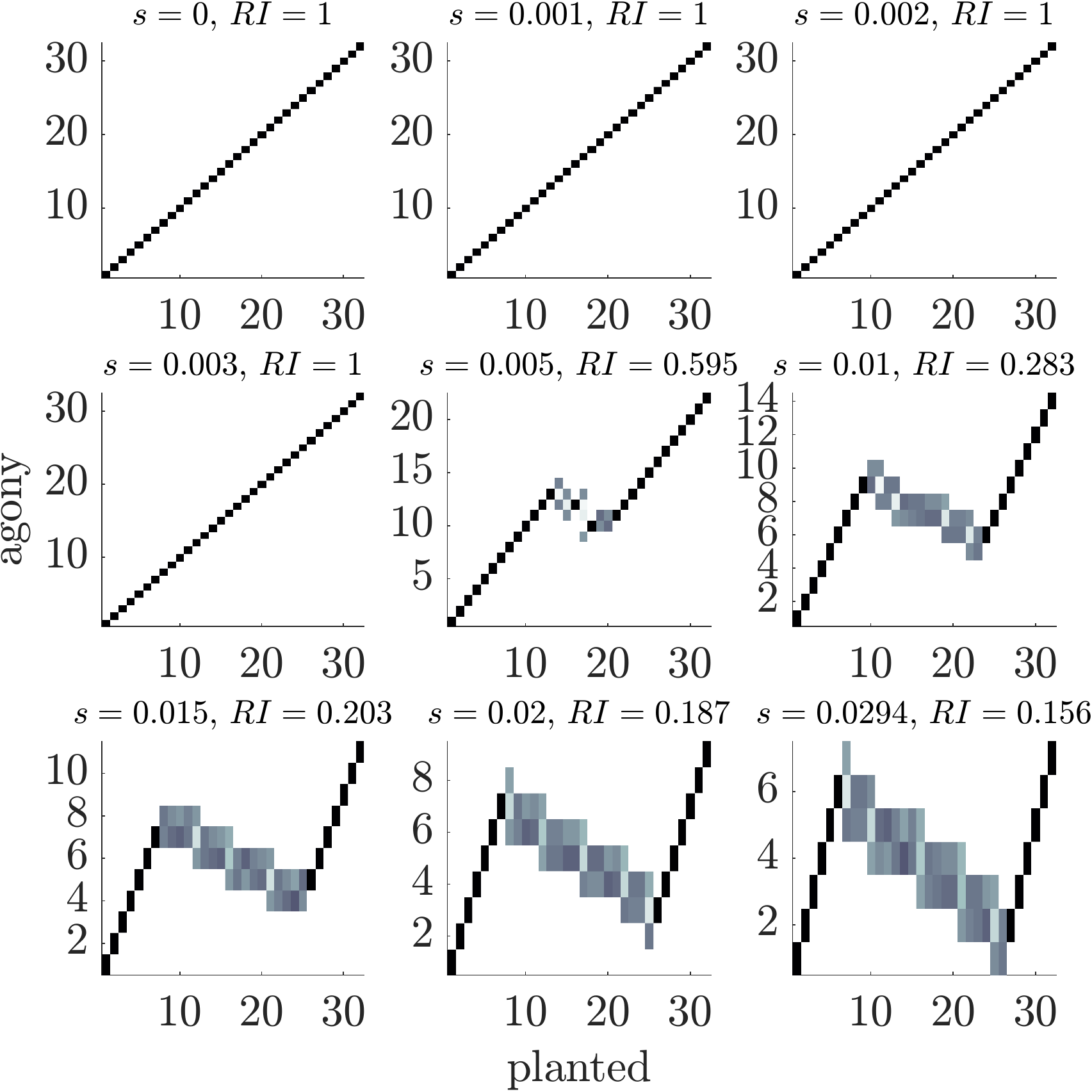}
\caption{Heat maps comparing the ranking inferred using {\it agony} with the planted ranking for military-like hierarchy. In each panel a square in position $(i,j)$ contains the number of nodes that belong to class $i$ in the planted rank and are placed in class $j$ by agony: the darker the color, the higher the number. The parameters are $p=0.5,\, q=0,\,R=32$, $s$ varies in $[0,s_{\max}]$, with $s_{\max}=0.0294$, and each plot refers to a single realisation of the ensemble.}
\label{fig:hmq0}
\end{figure}

For the military-like hierarchy things are more complicated. \cref{fig:hmq0} shows the heat map of the classes for $p=0.5$ and nine values of $s$. With these parameters our formulas give $s_i=0.00280$ and $s_1=0.00284$. We see that for strong hierarchical structures (small $s$) {\it agony} recovers well the classes. However when $s$ increases a {\it partial} inversion of the hierarchy is observed and only for large $s$ we recover the fully inverted ranking we studied in the previous Section. Thus simulations show that the latter is not always the optimal ranking but rather there are partially inverted rankings with a larger hierarchy. The purpose of the above analysis on the military-like hierarchy is to show that there exist values of the parameters for which the planted ranking is not optimal and to demonstrate that partial inversion can outperform the planted one. Moreover the partial inversion is observed for $s=0.002<s_1$, hence our computations provide a upper bound of the true resolution threshold. 

\begin{figure}[t]
\center
\includegraphics[scale=0.6]{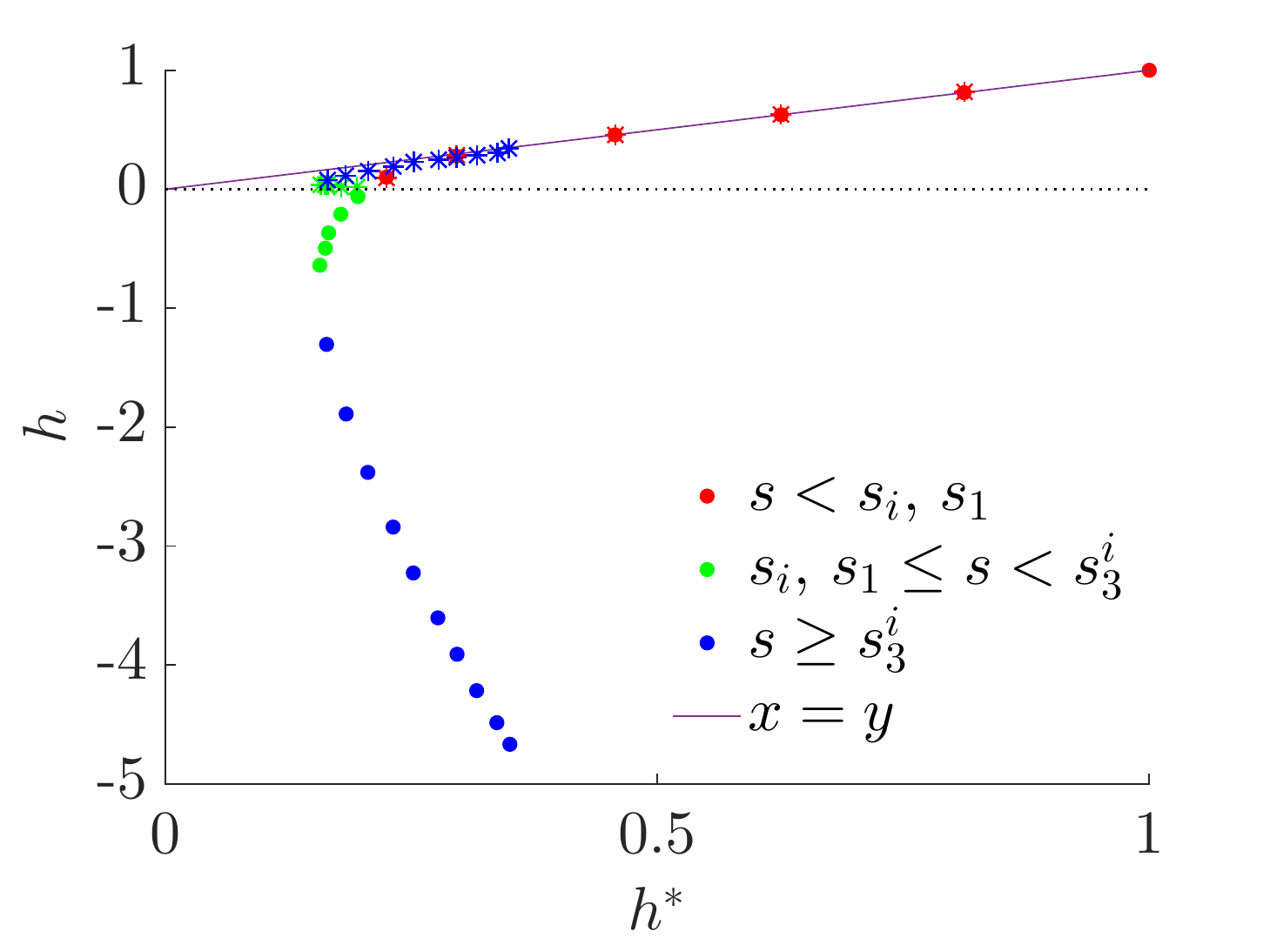}
\caption{Comparison of hierarchies of military-like RSBMs. The parameters are $p=0.5,\, q=0,\,R=32$, $s$ varies in $[0,s_{\max}]$, with $s_{\max}=0.0294$, and each point refers to a single realisation of the ensemble. The circles represent the pairs $(h_1^*,h_1^{(p)})$, i.e. the optimal hierarchy $h_1^*$ computed with {\it agony} and the one of the planted hierarchy $h_1^{(p)}$. The stars represent $(h_1^*,\bar{h}_1^*)$ where $\bar{h}_1^*$ is the theoretical hierarchy with the parameters of the SBM estimated via \emph{GraphTool}. 
}
\label{fig:hhq0}
\end{figure}

\cref{fig:hhq0} shows, similarly to \cref{fig:hh}, the scatter plot of the true value of $h_1^*$ computed via \emph{agony} on the simulated graphs  against the hierarchy of the planted rank $h_1^{(p)}$ (circles), and against $\bar{h}^*_1$, the hierarchy computed with Eq. \eqref{eq:hstar-d1} using the coefficients of the SBM estimated from the sample graph with \emph{GraphTool}. The main message of the figure is that, despite the fact the symmetrically inverted ranking is not the optimal one according to numerical simulations, its value of hierarchy is very close to the one of the optimal ranking, while the planted one strongly mis-estimates the value of $h$. Thus our computation in the previous Section can be used to reliably estimate the hierarchy of a military-like ranking. This is obviously a partial answer and analytical calculations of the hierarchy of partially inverted rankings are left for a future study.

\section{Beyond the resolution limit: Iterated agony}\label{sec:iter}

In the previous Sections we have shown theoretically and numerically that inference of ranking hierarchies based on agony suffers from significant resolution limit. In twitter-like hierarchies, the identified classes are merging of adjacent classes and thus small classes are not identified.  In military-like hierarchies inversions start to play a significant role.  

An heuristic method to overcome this problem is to iterate the application of agony. As done with modularity, one can apply agony to each class found in the first iteration of the algorithm, in order to find subclasses. In principle one could continue to iterate, even if the fact that agony finds two classes in an Erd\"os-Renyi graph suggests a careful design of the stopping criterion. The purpose of this Section is not to propose a full criterion for the improvement of agony via iteration, but to show that indeed improvement is possible, both considering model graphs and real networks.

\begin{figure}[t]
\center
\includegraphics[scale=0.6]{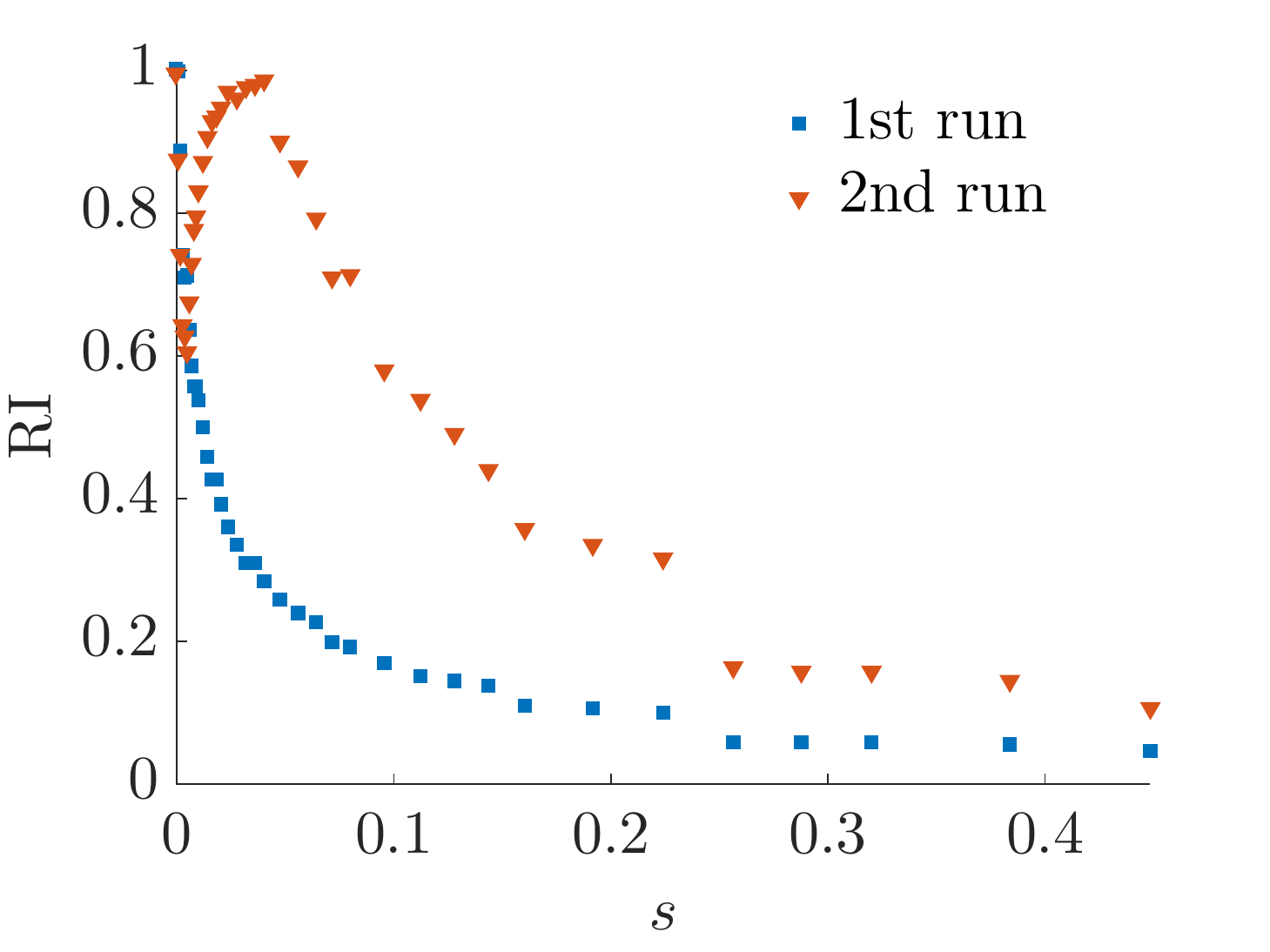}
\caption{Comparison of the Rand Index between the planted ranking and one (blue squares) or two (orange triangles) iterations of {\it agony}. Data refers to simulation of twitter-like HSBM with parameters $p=q=0.5,\,R=32,\,s\in [0,s_{\max}]$, with $s_{\max}=0.448$, and each point refers to a single realisation of the ensemble.}
\label{fig:hiter}
\end{figure}

We first consider the model graphs with twitter-like hierarchy we presented in the previous Section. Figure \ref{fig:hiter} shows the RI between the planted ranking and the one inferred with one (as in the previous Section) and two iterations of {\it agony} with $d=1$. For small values of $s$ the second iteration does not improve the inference because one iteration already recovers the planted structure. For larger values of $s$, i.e. weaker structures, the second iteration dramatically outperforms the result of the first one, indicating that iterated applications of {\it agony} can significantly improve the hierarchies detection. Figure \ref{tab:trial-sum} shows some details of the obtained results. It is worth noticing that the value of $h$ after the second run is actually smaller than the one from the first run, despite the fact that the RI follows the opposite pattern. This is expected since {\it agony} finds the optimal value of $h$, while the RI looks at the similarity with the planted ranking.
A closer look to the results of the two iterations (see Tables \ref{tab:c2345} and \ref{tab:c6789} in Appendix \ref{ap:res}) highlights that high number of classes after the second iteration and high hierarchy in each subclass are associated to the cases for which there is no significant improvement in the RI, hence a successful routine would rely on the control of these two quantities to 
stop the iterations.

\begin{table}[ht]
\center
\caption{Simulated graphs, output of the two runs of agony}\label{tab:trial-sum}
\begin{tabular}{c|cccccc}
&\multicolumn{3}{c}{1st run}&\multicolumn{3}{c}{2nd run}
\\
$s$&$h^*$&RI&$R$&$h$&RI&$R'$\\\hline
0&1&1&32&$>0.99$&$>0.99$&32\\
0.001&0.98&$>0.99$&34&0.93&0.87&97\\
0.002&0.95&0.89&29&0.81&0.74&128\\
0.005&0.91&0.71&20&0.51&0.60&160\\
0.01&0.85&0.54&14&0.51&0.83&102\\
0.048&0.62&0.26&7&-0.14&0.90&40\\
0.112&0.41&0.15&4&-0.16&0.54&17\\
0.224&0.20&0.10&3&-0.20&0.31&9\\
0.448&0.03&0.05&2&-0.19&0.11&4\\
\end{tabular}

\end{table}

We now show that the same phenomenon is relevant also for real networks. We investigate four datasets from SNAP, Stanford Network Analysis Platform \cite{snapnets}, which were also used in \cite{tatti2017tiers}\footnote{Note that these datasets have been updated since they have been used in \cite{tatti2017tiers} so our results are slightly different.}.

The networks are quite different in size (from a minimum of 7K nodes to almost 400K nodes) but they are all quite sparse. 
\begin{itemize}
\item {\bf Wiki vote.}
The network contains all the Wikipedia voting data from the inception of Wikipedia till January 2008. Nodes in the network represent Wikipedia users and a directed edge from node $i$ to node $j$ represents that user $i$ voted for user $j$.
\item {\bf Higgs Reply.}
The network contains replies to existing tweets: nodes are users and $i$ is linked to $j$ if $i$ replied to a $j$'s tweet.
\item {\bf Higgs mention. }
Similar to the previous case, here links represent mentions: a link from $i$ to $j$ means that user $i$ mentioned user $j$.
\item {\bf Amazon.} Network was collected by crawling the Amazon website. It is based on Customers Who Bought This Item Also Bought feature of the Amazon website. If a product $i$ is frequently co-purchased with product $j$, the graph contains a directed edge from $i$ to $j$.
\end{itemize}

Table \ref{tab:sum} reports some properties of the networks alongside the output of one and two iterations of the {\it agony} algorithm. Specifically, for each network the table contains: the number of nodes $N$, the density ($\frac{m}{N(N+1)}$, where $m$ is the number of edges), the percentage of nodes in the largest strongly connected component (\emph{SCC}), the value of $h_1^*$, the number of classes inferred in the first run ($R$) and the total number of classes after the second run ($R'$) of {\it agony}.

\begin{table}[ht]
\center
\caption{Networks summary. SCC is the percentage of nodes in the largest strongly connected component, $h_1^*$ is the hierarchy of the ranking obtained with one iteration of {\it agony}, $R$ is the number of classes in the globally optimal ranking, and $R'$ is the number of classes after two iterations of {\it agony}.}\label{tab:sum}
\begin{tabular}{crcccccc}
network & nodes & density & SCC &$h_1^*$ & $R$ &$R'$\\ 
\hline 
Wikivote & $7,115 $&$ 2*10^{-3}$ &$ 18\%$&$0.83$ &$ 12$&$49$ \\ 
HiggsReply & $38,918$ &$ 2*10^{-5}$ &$ 0.8\%$&$0.82$ & $13$&$27$\\  
HiggsMention & $116,408$ & $1*10^{-5}$ &$ 1\%$ &$0.89$&$ 20$&$59$ \\ 
Amazon & $403,394$ & $2*10^{-5}$ &$ 98\% $&$0.42$& $17 $&$69$
\end{tabular}
\end{table}

It is clear that the second application the algorithm to the classes detected in the first iteration increases significantly the number of classes, suggesting that the classes identified in the first iteration could be aggregation of smaller classes. In Table \ref{tab:bench-h} in Appendix \ref{ap:res} we report more details on the classes identified in the iteration and on the subclasses identified by the second iteration.

Since \emph{agony} penalises links among nodes in the same class, the subgraphs in some cases have no links (those with $*$ in Table \ref{tab:bench-h} in Appendix \ref{ap:res}). Notice this would be the case for any class in a DAG. Thus, a low value of $h$ in each class and a number of sub classes larger than $2$ indicate a non trivial and not completely resolved structure of the class.

\section{Conclusion}\label{sec:concl}

In this paper we have studied the inference of hierarchical structures in directed networks by introducing an ensemble of random graphs, termed the Hierarchical Stochastic Block Model, and studying how  agonies, penalising links contrary to the hierarchy, are able to identify the planted ranking.  

Using symmetry arguments we have explored ranking alternative to the planted one and obtained from it by merging, splitting or inverting its classes. We have shown that when the hierarchy is not strong enough some of these alternative rankings of nodes have a value of the hierarchy larger than the planted one. This demonstrates that (generalised) agonies have a resolution limit, being unable to detect small classes in large networks. This is somewhat similar to the well known resolution limit of modularity in community detection. In some cases we have strong numerical indications that the proposed alternative rankings, are actually close to the global optimal one. Finally we have shown that in these cases the iterated application of agony can lead to significant improvement of the hierarchy detection.

There are several directions along which our work can be extended. First, we have investigated in detail the case of uniform cardinality of the classes, even if our formulae can be used to study more complex structures, such as a pyramidal hierarchy with a small top class and larger bottom classes. With a careful choice of the sizes one might be able to maintain analytical tractability, however the study of these structures are left for future investigation. The second direction is to consider, at least theoretically, other values of $d$ (or other agony functions). We have shown some results indicating that the resolution threshold depends on $d$, however numerical simulations cannot be performed because of the lack of heuristic methods for optimisation of agony with $d\ne 0,1$. Finally, other methods to identify ranking hierarchies could be investigated, for example suitably modifying the agony function or by considering optimisations for a set of functions. 

We leave these extensions for future work and we are confident that the results will be of interest in the general problem of hierarchy detection in networks.
\appendix
\section{Detailed proofs}\label{ap}
In this section we present details and extendes formulae for the propositions in \cref{sec:optimalHier}.

To start, we consider the values of agony for general $d$ depending on the choice of the alternative rankings.
\begin{itemize}
\item {\bf No inversion and splitting.} When $b<0$, each class is divided into $2^{-b}$ classes. As for the affinity matrix, the only part affected by the change in the ranking is the one above the diagonal, which has no impact on the computation of $\mathbb{E}[A_d(G,r^{(b)})]$. Hence one has
\begin{align}
\mathbb{E}[A_d(G,r^{(b)})]&=s\left(\frac{N}{R}2^{b}\right)^2\sum_{k=0}^{2^{a-b}-1}(k+1)^d(2^{a-b}-k)\,.\label{eq:ad}
\end{align}

\item {\bf No inversion and merging.} When $b\geq0$, for any pair $(i,j)$ it holds:
\begin{equation}
\mathbb{E}[m_{ij}]=\left(\frac{N}{R}\right)^2\begin{cases}
2^{2b}s&j<i\\
(2^b-1)p+2^{b-1}(2^b+1)s+(2^{b-1}-1)(2^b-1)q& j=i\\
p+(2^{2b}-1)q& j=i+1\\
2^{2b}q& j>i+1\,,
\end{cases}\label{eq:merge}
\end{equation}
which gives
\begin{align*}
\mathbb{E}[A_d(G,r^{(b)})]=&s\left(\frac{N}{R}2^{b}\right)^2\sum_{k=1}^{2^{a-b}-1}(k+1)^d(2^{a-b}-k)+\\
&+2^{a-b}\left((2^b-1)p+2^{b-1}(2^b+1)s+(2^{b-1}-1)(2^b-1)q\right)
\end{align*}

\item {\bf Inversion and merging.} When $b\geq0$ the expression for agony of the inverted ranking becomes
\begin{align*}
\mathbb{E}[A_d(G,r^{(i,b)})]=&2^{2 b}\left(\frac{N}{R}\right)^2 q \sum _{k=2}^{2^{a-b}-1} (k+1)^d \left(2^{a-b}-k\right)+\\
&+2^d \left(\frac{N}{R}\right)^2\left(2^{a-b}-1\right) \left(\left(2^{2 b}-1\right) q+p\right)+\\
&+2^{a-b} \left(\frac{N}{R}\right)^2\left(\left(2^b-1\right) p+\left(2^{b-1}-1\right)
   \left(2^b-1\right) q+2^{b-1} \left(2^b+1\right) s\right)
\end{align*}

\item {\bf Inversion and splitting}  When $b<0$ 
\begin{align*}
\mathbb{E}[A_d(G,r^{(i,b)})]=&\left(\frac{N}{R}2^{b}\right)^2\sum _{k=0}^{2^{-b}-1} (k+1)^d \left(2^a
   \left(2^{-b}-k\right)s+\left(2^a-1\right) k p\right)+\\
&+\left(\frac{N}{R}2^{b}\right)^2 \sum _{k=0}^{2^{-b}-1} \left(k+1+2^{-b}\right)^d \left(\left(2^a-1\right) \left(2^{-b}-k\right)p+\left(2^a-2\right) k q\right)+\\
&+\left(\frac{N}{R}2^{b}\right)^2q\sum _{k=0}^{\left(2^a-2\right) 2^{-b}} \left(k+1+2^{1-b}\right)^d\left(\left(2^a-2\right) 2^{-b}-k\right)\,.
\end{align*}
\end{itemize}

Then, we present the proofs of the propositions.

\subsection{Proof of Proposition \ref{prop:d1}}
We explicitly show that in the $d=1$ case there exists critical values for $s$ at which the planted ranking ceases to maximize hierarchy both for Twitter-like and Military-like hierarchies. 

To determine the optimal number of classes we first treat $b$ as a continuous variable and compute the derivative oh $\bar{h}_1$ with respect to it. The unique critical point is denoted by $b^*$ and it is given by
\[b^*=\frac{1}{2}\log_2{\frac{2^{2a}s+6(q-p)}{3q-s}}\,.
\]
Note that it must hold
\[ 0\leq b\leq a\]
and we want to avoid the continuous relaxation at the boundaries so we consider the extreme values separately.

When $p\geq q>s$ (\emph{Twitter-like hierarchy}), we first notice that
\[
\frac{\partial \bar{h}_1}{\partial b}\lvert_{b=b*}<0
\]
Moreover, it holds  
\[\bar{h}_{1}({b=a-1})>\bar{h}_{1}({b=a})\,,\]
that is the trivial ranking is never better than that with two classes.

Moreover, we denote with $s_2$ the value of $s$ such that the rankings with two and three classes have the same value of hierarchy, i.e.
\[
\bar{h}_{1}\left({b=a-\log_23}\right) =\bar{h}_1({b=a-1})\,,\]
since for any fixed $b>0$, $\bar{h}_{1}$ is monotone decreasing with respect to $s$,
\[\bar{h}_{1}\left({b=a-\log_23}\right) <\bar{h}_{1}({b=a-1})\,\forall s\geq s_2\,.\]
Similarly, one can find the critical value $s_m$ such that the ranking with of $R-1$ classes shares the value of hierarchy with the planted one,
\[\bar{h}_{1}({b=0}) =\bar{h}_{1}\left({b=a-\log_2{(2^a-1)}}\right)\,.\]

Finally, we can combine the results to obtain the optimal number of classes for the direct ranking in the region $p\geq q>s$:
\begin{equation}
\tilde R^*=\begin{cases}R&
s\leq s_m \\
2^{a-b^*} & s_m<s< s_2\\
2 & s\geq s_2
\,,
\end{cases}\label{eq:rstar}
\end{equation} 
where
\begin{align}
s_m&=\frac{6 \left(2^a-1\right) p-3 \left(2^a-2\right) q}{2^a-4^a+8^a}\label{eq:smd1}\\
s_2&=\frac{3}{7}\frac{\left(4^a-12\right) q+12 p}{ 4^{a}}\notag
\end{align}
With a reasoning similar to the one carried before, one gets that when $p\geq q >s$ the optimal number of classes for the inverted ranking is such that 
\[ 1\leq \tilde R^*\leq 2\]
hence, 
\[h_1^{i,*}\leq 0,\, \forall \, p\geq q >s,\,\forall\,a\,.\]
One can conclude that the optimal ranking for the twitter-like hierarchy is the direct one with a number of classes which depends on $s$, according to \eqref{eq:rstar}.

When $q=0$ (\emph{Military-like hierarchy}), when it is defined, we have
\[\frac{\partial^2 \bar{h}_1}{\partial b^2}\lvert_{b=b*}>0\,,\]
so, to obtain the optimal directed ranking we only need to check the extreme values for $b$, i.e. $b=0,\,b=a$.
The optimal number of classes for the direct ranking is given by
\[ \tilde R^*=\begin{cases}R&
s\leq s_{m\lvert_{q=0}} \\
1 & \text{otherwise}\,,
\end{cases}
\]
where
\[s_{1}=\frac{6p}{2^a(1+2^a)}\,.\]

Then, one can consider the inverted ranking. 

It easy to verify that
\[\mathbb{E}[A_1(G,r^{(i,b)})]>\mathbb{E}[A_1(G,r^{(p)})],\,\forall\,b<0\,,\]
that is, also for the inverted ranking splitting is never optimal on average.\\

As for merging, the optimal choice for $b$ is given by
\[b^{i,*}=\frac{1}{2}log_2{\frac{2p}{s}}\,,\]
which is well defined when $s>\frac{2}{4^a}p$ and satisfies $\frac{a}{2}\leq b^{i,*}\leq a$.
The optimal number of classes fro the inverted ranking is given by
\[\tilde R^{i,*}=\begin{cases}
1&s\leq s_2^i\\
2& s_2^i<s\leq s_3^i\\
2^{a-b^{i,*}}& s>s_3^i
\end{cases}\,,
\]
where
\begin{align*}
s_2^i&=2^{2-2a}p\\
s_3^i&=3s_2^i\,.
\end{align*}
When $s\leq s_{1}$, the planted ranking is optimal and non zero and decreasing, and
\begin{equation}
s_2^i< s_{1}<s_3^i\,. \label{eq:inv}
\end{equation}
Denote by $s_i$ the value of $s$ such that
\[
\bar{h}^i_1({b=a-1})=\bar{h}_1({b=0})\,.
\]
One gets
\[s_i= \frac{12 p}{3\ 2^a+2^{2 a+1}-2}\,,\]
and when $s>s_i$ the optimal inverted ranking has a higher value of hierarchy than the planted, which is the optimal directed one.

Finally, one can write the expression for the estimate of the optimal value of $h$ in proposition \ref{prop:d1}.

For $p\geq q>s$,
\[
\bar{h}_1^*=\begin{cases}
 -\frac{\left(2^a-2\right) \left(-6 \left(2^a-1\right) q+2^a \left(2^a+2\right) s-6 p\right)}{6 \left(2^a (2 p-3 q+s)+4^a (q+s)-2 p+2 q\right)} & s\leq s_m
   \\
 \frac{3 \left(\left(4^a+2\right) q-2 p\right) \sqrt{\frac{4^a s-6 p+6 q}{3 q-s}}-2^{a+1} \left(4^a s-6 p+6 q\right)}{3 \sqrt{\frac{4^a s-6 p+6 q}{3 q-s}} \left(2^a (2 p-3
   q+s)+4^a (q+s)-2 p+2 q\right)} & s_m<s<s_2 \\
 \frac{4^a (q-s)+4 p-4 q}{2 \left(2^a (2 p-3 q+s)+4^a (q+s)-2 p+2 q\right)} & s\geq s_2\,.
\end{cases}
\]
When $q=0$,
\[
\bar{h}_1^*=
\begin{cases}
 \frac{2^a (6 p+s)-8^a s-6 p}{6 \left(2^a-1\right) p+3\ 2^a \left(2^a+1\right) s} & s\leq  s_i \\
 \frac{4^a s-4 p}{2 \left(2^a (2 p+s)+4^a s-2 p\right)} & s_i<s\leq s_3^i \\
 \frac{-2^{a+\frac{3}{2}} s \sqrt{\frac{p}{s}}+4^a s+2 p}{2^a (2 p+s)+4^a s-2 p} &  s>s_3^i\,.
\end{cases}
\]

\subsection{Proof of Proposition \ref{p:d0}}
We here proceed to show that in the $d=0$ case (FAS), both for Twitter-like and Military-like hierarchies, agony is minimized by the ranking where nodes are partitioned in singletons.
When $b>0$, the derivative of $h$ with respect to $b$ is negative hence the planted ranking is better that any other with a fewer number of classes. Instead, when $b<0$ one has
\[\mathbb{E}[A_0(G,r^{(b)})]=s(2^a+2^b)\left(\frac{N}{R}\right)^2\,
\]
which implies 
\[\mathbb{E}[A_0(G,r^{(b)})]<\mathbb{E}[A_0(G,r^{(p)})],\,\forall\,b<0\,,\]
and
\[\frac{\partial \bar{h}_0}{\partial b}=-\frac{2^{a+b-1}}{m}s<0\quad\forall\,b<0\]
So the optimal ranking is obtained for the limit value of $b$
\[ b^*=-\log_2{\frac{N}{R}},\,\quad \tilde R^*
=N\,.\]

Similar computations give that any inverted ranking (i.e $\forall\, b $) has never a higher value of hierarchy than the the ranking we just discussed.

One get the formula in proposition \ref{p:d0}
\[h_0^*=1-\frac{2^{2a} (N+1)s }{ \left(2^{2a} (q+s)+2^a (2 p-3 q+s)-2 p+2 q\right)N}
\]

\subsection{Proof of Proposition \ref{p:d2}}
For the case $d=2$ one can follow the same procedure we showed for $d=1$ and find the critical values for resolution threshold.

When $p\geq q>s$, the optimal number of classes is given by
\[ \tilde R_2^*=\begin{cases}R& s\leq s_{2,m} \\
2^{a-b_2^*}& s_{2,m}\leq s\leq s_{2,1}\\
1 & s\geq s_{2,1}\,,
\end{cases}
\]
where
\begin{align}
b_2^*=&\log_2 (\frac{2 \sqrt[3]{2} \left(2^{2a} s-3 p+3 q\right)}{\sqrt[3]{\beta+3^5\ 2^{3a} q^2 s-3^4\ 2^{3 a+2} q
   s^2+3^3\ 2^{3 a+2} s^3}}+\label{eq:b2star}\\
   &+\frac{\sqrt[3]{\frac{1}{3} \beta+2^4\ 2^{3a} q^2 s-3^3\ 2^{3 a+2} q s^2+3^2\ 2^{3 a+2}
   s^3}}{\sqrt[3]{2\,3^2} (3 q-2 s)})\,,\notag\\
   \beta=&\sqrt{3^6\ 2^{6a} s^2 (3 q-2 s)^4-2^5\,3^3 (3 q-2 s)^3 \left(4^a s-3 p+3 q\right)^3}\,.\notag
\end{align}
is the unique zero of the first order derivative of $\bar{h}_2$ with respect to $b$, and
\begin{align*}
s_{2,m}&=\frac{6 \left(2^{1-a} (q-p)+2 p-q\right)}{-3\ 2^a+2^{3 a+1}+4^a+4}\\
s_{2,1}&=\frac{2^{2 a} q+4 p-4 q}{3\,2^{2 a}}\,
\end{align*}
with $s_{2,1}$ being the value of $s$ such that 
\[
\bar{h}_{2}(b=a-1)=\bar{h}_{2}(b=a)=0\,.
\]

When $q=0$, the planted ranking is optimal and gives positive $\bar{h}_2$ when $s<s_{2,1}^0$, where
\[
s_{2,1}^0=\frac{3\ 2^{2-a} p}{5\ 2^a+4^a+4}\,.
\]
For the inverted ranking instead one can compute the optimal choice for the number of classes, that is
\[\tilde R_2^{i,*}=\begin{cases}
 a & s\leq s_{2,2}^i \\
 a-1 & s_{2,2}^i p<s<s_{2,3}^i\\
 \frac{\log \left(\frac{6 p}{s}\right)}{\log (4)} & s>s_{2,3}^i\,,
\end{cases}
\]
where
\[
b_2^{i,*}=\frac{\log \left(\frac{6 p}{s}\right)}{2 \log (2)}\,,
\]
and
\begin{align*}
s_{2,2}^i&=\frac{12}{2^{2a}} p\\
s_{2,3}^i&=3 s_{2,2}^i\,.
\end{align*}
For any choice of $p$ and $a$, it holds
\[
s_{2,1}< s_{2,2}^i\,,
\]
so the inverted ranking is optimal for $s>s_{2,2}^i$.

\section{Numerical results}\label{ap:res}
In this section we provide additional results from the empirical applications.

For each network and for each class, the table contains the size of the class, $n_i$, as a percentage of the total number of nodes), the value of $h$,  and the number of sub-classes inferred (R).

\begin{table}[ht]
\center
\caption{Simulated graphs, details for classes}\label{tab:c2345}
\begin{tabular}{c|cccccccccccc}
&\multicolumn{3}{c}{$s=0.001$}&\multicolumn{3}{c}{$s=0.002$}&\multicolumn{3}{c}{$s=0.005$} &\multicolumn{3}{c}{$s=0.01$}\\
cl.&$n_i(\%)$&$h^*$&$R$&$n_i(\%)$&$h^*$&$R$&$n_i(\%)$&$h^*$&$R$&$n_i(\%)$&$h^*$&$R$\\
\hline
1&$<$0.01&1*&1	&0.03&1&3			&0.03&1&11			&0.03&0.95&12\\
2&0.03&1&3		&0.03&1&3			&0.03&1&8			&0.06&0.93&5\\
3&0.03&1&4		&0.03&0.88&3		&0.03&1&7			&0.06&0.93&6\\
4&0.03&1&3		&0.03&1&4			&0.03&0.98&7		&0.09&0.93&5\\
5&0.03&1&3		&0.03&1&3			&0.06&0.97&7		&0.09&0.94&5\\
6&0.03&1&3		&0.03&1&4			&0.06&0.97&7		&0.095&0.95&5\\
7&0.03&1&3		&0.03&1&5			&0.06&0.97&10		&0.09&0.94&6\\
8&0.03&1&2		&0.03&1&5			&0.06&0.97&9		&0.09&0.93&6\\
9&0.03&1&3		&0.03&1&3			&0.06&0.96&8		&0.09&0.94&7\\
10&0.03&1&2		&0.03&1&4			&0.06&0.97&8		&0.09&0.94&6\\
11&0.03&1&3		&0.03&1&4			&0.06&0.97&11		&0.06&0.93&6\\
12&0.03&1&2		&0.045&0.99&5		&0.06&0.97&6		&0.06&0.94&7\\
13&0.03&1&4		&0.05&0.99&6		&0.06&0.97&10		&0.03&0.87&10\\
14&0.03&1&2		&0.05&0.99&8		&0.06&0.97&8		&0.03&0.91&16\\
15&0.03&1&3		&0.05&0.98&7		&0.06&0.97&7\\
16&0.03&1&2		&0.05&0.99&6		&0.06&0.97&7\\
17&0.03&1&3		&0.04&0.99&5		&0.03&0.94&7\\
18&0.03&1&3		&0.03&1&5			&0.03&1&6\\
19&0.03&1&3		&0.03&1&4			&0.03&1&9\\
20&0.03&1&3		&0.03&1&5			&0.03&1&7\\
21&0.03&1&3		&0.03&1&4\\
22&0.03&1&3		&0.03&1&4\\
23&0.03&1&3		&0.03&1&4\\
24&0.03&1&3		&0.03&1&3\\
25&0.03&1&5		&0.03&1&4\\
26&0.03&1&3		&0.03&1&5\\
27&0.03&1&3		&0.03&1&4\\
28&0.03&1&3		&0.03&1&3\\
29&0.03&1&2		&0.03&1&5\\
30&0.03&1&3\\
31&0.03&1&4\\
32&0.03&1&3\\
33&0.03&1&3\\
34&$<$0.01&1*&1\\
\end{tabular}
\end{table}

\begin{table}[ht]
\center
\caption{Simulated graphs, details for classes (ctd)}\label{tab:c6789}
\begin{tabular}{c|cccccccccccc}
&\multicolumn{3}{c}{$s= 0.048$}&\multicolumn{3}{c}{$s=0.112$} &\multicolumn{3}{c}{$s=0.224$}&\multicolumn{3}{c}{$s= 0.448$}\\
cl.&$n_i(\%)$&$h$&$R$&$n_i(\%)$&$h$&$R$&$n_i(\%)$&$h$&$R$&$n_i(\%)$&$h$&$R$\\
\hline
1&0.06&0.71&4		&0.18&0.44&4		&0.27&0.21&3		&0.51&0.03&2\\
2&0.16&0.70&6		&0.32&0.42&5		&0.48&0.20&3		&0.49&0.03&2\\
3&0.22&0.67&7		&0.32&0.42&4		&0.26&0.21&3\\
4&0.22&0.67&7		&0.18&0.44&4\\
5&0.19&0.69&7\\
6&0.12&0.72&4\\
7&0.03&0.35&5\\
\end{tabular}
\end{table}

\begin{table}[ht]
\center
\caption{Real networks: details for classes}\label{tab:bench-h}
\begin{tabular}{c|cccccccccccc}
&\multicolumn{3}{c}{Wikivote}&\multicolumn{3}{c}{HiggsReply}&\multicolumn{3}{c}{HiggsMention} &\multicolumn{3}{c}{Amazon}\\
cl.&$n_i(\%)$&$h^*$&$R$&$n_i(\%)$&$h^*$&$R$&$n_i(\%)$&$h^*$&$R$&$n_i(\%)$&$h^*$&$R$\\
\hline
1 		&$0.67$&$1$*&$1$		&$0.60$&$0.03$&$2$	&$0.77$&$0.13$&$2$	&$0.02$&$<0.01$&$3$\\
2 		&$0.01$&$0$	&$1$		&$0.31$&$0.34$&$2$	&$0.16$&$0.78$&$4$	&$0.03$&$0.01$&$3$\\
3 		&$<0.01$&$0.24$&$3$		&$0.04$&$0.64$&$2$	&$0.03$&$0.78$&$3$	&$0.10$&$0.01$&$3$\\
4 		&$0.01$&$0.38$&$6$		&$0.01$&$1$&$2$		&$0.01$&$0.80$&$2$	&$0.20$&$0.01$&$5$\\
5 		&$0.02$&$0.26$&$5$		&$<0.01$&$0.50$&$2$	&$<0.01$&$0.81$&$2$	&$0.25$&$0.05$&$6$\\
6 		&$0.04$&$0.23$&$6$		&$<0.01$&$1$&$2$	&$<0.01$&$0.85$&$2$	&$0.20$&$0.08$&$6$\\
7 		&$0.06$&$0.20$&$5$		&$<0.01$&$1$&$3$	&$<0.01$&$0.53$&$2$	&$0.11$&$0.08$&$6$\\
8 		&$0.09$&$0.32$&$8$		&$<0.01$&$0.67$&$5$	&$<0.01$&$0.65$&$3$	&$0.06$&$0.07$&$5$\\
9 		&$0.08$&$0.72$&$10$		&$0.01$&$0.24$&$2$	&$<0.01$&$0.66$&$4$	&$0.02$&$0.06$&$5$\\
10		&$0.04$&$1$&$2$			&$<0.01$&$0.83$&$2$	&$<0.01$&$0.55$&$4$	&$0.01$&$0.05$&$5$\\
11		&$<0.01$&$0$&$1$		&$<0.01$&$1$*&$1$	&$<0.01$&$0.59$&$7$	&$<0.01$&$0.06$&$4$\\
12		&$<0.01$&$1$*&$1$		&$<0.01$&$1$*&$1$	&$<0.01$&$0.46$&$6$	&$<0.01$&$0.05$&$4$\\
13		&&&						&$<0.01$&$1$*&$1$	&$<0.01$&$0.60$&$6$	&$<0.01$&$0.04$&$3$\\
14		&&&						&&&					&$<0.01$&$0.66$&$6$	&$<0.01$&$0.05$&$4$\\
15 	&&&						&&&					&$<0.01$&$0.82$&$1$	&$<0.01$&$0.07$&$3$\\
16 	&&&						&&&					&$<0.01$&$0.69$&$1$	&$<0.01$&$0.05$&$3$\\
17 	&&&						&&&					&$<0.01$&$1$*&$1$	&$<0.01$&$0$&$1$\\
18 	&&&						&&&					&$<0.01$&$1$*&$1$	&&		\\
19 	&&&						&&&					&$<0.01$&$1$*&$1$	&&		\\
20  	&&&						&&&					&$<0.01$&$1$*&$1$	&&		\\
\end{tabular}

* empty
\end{table}

\section*{Acknowledgment}
PB acknowledges support from FET Project DOLFINS nr. 640772 and FET IP Project MULTIPLEX nr. 317532;
FL acknowledges support by the European Community's H2020 Program under the scheme INFRAIA-1- 2014-2015: Research Infrastructures, grant agreement no. 654024 SoBigData: Social Mining \& Big Data Ecosystem.

\bibliographystyle{acm}
\bibliography{draft_revised/Agony}

\end{document}